\documentclass[10pt,pra,aps,showpacs,twocolumn,unsortedaddress,longbibliography]{revtex4-1}

\usepackage{graphicx,bm}
\usepackage{amsmath}
\usepackage{amssymb}
\usepackage{epsfig}
\usepackage{epsf}
\usepackage{verbatim}
\usepackage[usenames]{color}
\usepackage[colorlinks=true,linkcolor=blue]{hyperref}

\newcommand{\BdotSigma}{\vec B \cdot \vec \sigma }
\newcommand{\TildeBdotSigma}{\vec {\tilde B} \cdot \vec {\tilde \sigma}}
\newcommand{\bs}{\bm}

\def\be#1\ee{\begin{equation}#1\end{equation}}
\def\ba#1\ea{\begin{align}#1\end{align}}
\def\bsa#1\esa{\begin{subequations}\begin{align}#1\end{align}\end{subequations}}
\newcommand{\rabi}{\Omega}

\begin{document}

\title{Phase space curvature in spin-orbit coupled ultracold atom systems}

\author{J. Armaitis}
\email{jogundas.armaitis@tfai.vu.lt}
\author{J. Ruseckas}
\email{julius.ruseckas@tfai.vu.lt}
\author{E. Anisimovas}
\email{egidijus.anisimovas@ff.vu.lt}
\affiliation{Institute of Theoretical Physics and Astronomy, Vilnius University,
Saul\.etekio Ave.~3, LT-10222 Vilnius, Lithuania}

\date{\today{}}

\begin{abstract}
We consider a system with spin-orbit coupling and derive equations of motion   
which include the effects of Berry curvatures.                                 
We apply these equations to investigate the dynamics of particles with equal   
Rashba-Dresselhaus spin-orbit coupling in one dimension.                       
In our derivation, the adiabatic transformation is performed first and leads   
to quantum Heisenberg equations of motion for momentum and position            
operators. These equations explicitly contain position-space, momentum-space,  
and phase-space Berry curvature terms. Subsequently, we perform the            
semiclassical approximation, and obtain the semiclassical equations of motion. 
Taking the low-Berry-curvature limit results in equations that can be          
directly compared to previous results for the motion of wavepackets.           
Finally, we show that in the semiclassical regime, the effective mass          
of the equal Rashba-Dresselhaus spin-orbit coupled system can be viewed        
as a direct effect of the phase-space Berry curvature.
\end{abstract}

\maketitle

\section{Introduction}

The geometrical concept of curvature has found multiple applications in various
branches of physics \cite{Frankel2004}, including the general theory of
relativity \cite{Einstein1915}, gauge theories in particle physics
\cite{YangMills1954}, and most recently condensed-matter physics in the guise
of Berry curvatures \cite{XiaoEtAl2010}. 
In simple terms, position-space Berry curvature can be understood 
as a result of magnetization texture in real space, 
while momentum-space Berry curvature requires 
spin-orbit coupling (SOC) \cite{XiaoEtAl2010, FujitaEtAl2011}.
Here, SOC is understood in the broad sense, i.e., as linking the velocity
to some quantized internal charcteristic of the particle.

Even though SOC arises naturally in crystals that lack an inversion
center, that is not the case in ultra-cold atom systems
\cite{ManchonEtAl2015}. There, the coupling between
the motion of each neutral atom and its hyperfine spin \cite{StuhlEtAl2015}
(or other degrees of freedom \cite{LiviEtAl2016, WallEtAl2016})
has to be artificially engineered \cite{DalibardEtAl2011}.
Recently, this field has seen considerable progress
\cite{GalitskiSpielm2013, GoldmanEtAl2014, Zhai2015}, and some of the
proposed spin-orbit coupling schemes have been experimentally realized.
In particular, one-dimensional equal Rashba-Dresselhaus 
\cite{Rashba1960, Dresselhaus1955} SOC was 
implemented several years ago \cite{LinEtAl2011a} and
has received a substantial amount of attention \cite{LiEtAl2015}. Furthermore,
there has been promising experimental progress
in engineering two-dimensional Rashba SOC
\cite{CampbellSpielman2016, HuangEtAl2016}, while three-dimensional
Weyl SOC remains an active theoretical research direction
\cite{AndersonEtAl2012, AndersonEtAl2013, DubcekEtAl2015, ArmaitisEtAl2017}.
Most of this work is concentrated on utilizing the internal states
of the atom and transitions between them with no spatial dependence.
However, other means, such as spatial degrees of freedom and periodic
driving of the system can also be efficiently exploited for similar purposes
\cite{StruckEtAl2012, StruckEtAl2014, Eckardt2016}.
For example, it is possible to achieve strong effective magnetic field 
in optical lattices and thus simulate various condensed-matter
Hamiltonians \cite{StruckEtAl2011, LewensteinEtAl2012, 
StruckEtAl2013, AidelsburgerEtAl2013, MiyakeEtAl2013, GoldmanEtAl2016}.
According to the concept of synthetic dimensions \cite{CeliEtAl2014}, internal 
degrees of freedom can be used to emulate additional spatial directions
\cite{CeliEtAl2014, ManciniEtAl2015, PriceEtAl2015a, StuhlEtAl2015, 
SuszalskiZakrzewski2016, AnisimovasEtAl2016, LiviEtAl2016}.
In this context, coupling between spatial and internal degrees of freedom was
demonstrated using both hyperfine \cite{StuhlEtAl2015} and long-lived
electronic \cite{LiviEtAl2016, WallEtAl2016} states.  Central to understanding
all of these advances is the notion of the Berry phase \cite{Berry1984}.

\begin{figure}[t]
\begin{center}
\includegraphics[width=\columnwidth]{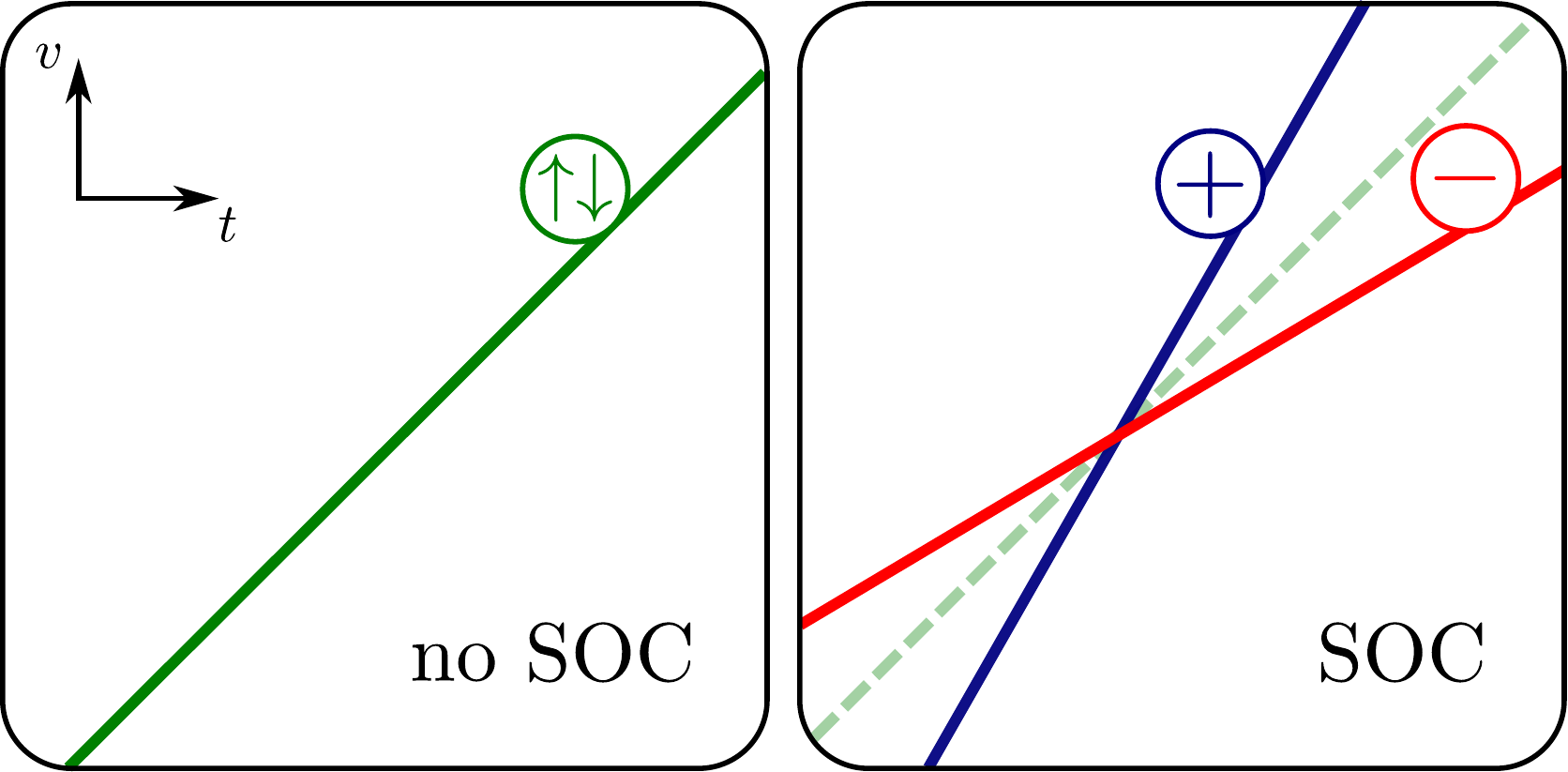}
\caption{In the absence of spin-orbit coupling both
spin species ($\uparrow$ and $\downarrow$) respond to a 
spin-independent linear potential in the same
way, since their masses are equal (left). When the spin-orbit
coupling is turned on, adiabatic motion occurs in dispersion branches
labelled by $+$ and $-$. The effective mass in the two branches
is different, resulting in a different response to the
same potential (right). This effect can be explained by the phase-space
Berry curvature in the semiclassical regime (see Sec.~\ref{sec:spiral}
for more details).}
\label{fig1}
\end{center}
\end{figure}

The Berry phase, as well as Berry curvatures in real and momentum spaces have
been thoroughly discussed in literature in various contexts, see
\cite{XiaoEtAl2010, GoldmanEtAl2014} and references therein. However, up to now
considerably less attention has been paid to phase-space Berry curvatures,
especially outside the solid-state physics community.  
It was only recently realized that this phase-space Berry curvature can lead to
an alternation of density of states \cite{FreimuthEtAl2013,
Bamler2016}.  This change in the density of states might in turn be used to
detect topological objects, such as skyrmions, by a mundane electrical
measurement.  This turns out to be particularly relevant to solid-state
materials, where both spin-orbit coupling and magnetization textures are
present \cite{HannekenEtAl2015, HamamotoEtAl2016}.

Theoretical progress concerning phase-space Berry curvature has been
mostly concentrated on lattice systems in the semiclassical approximation. 
In an early publication~\cite{SundaramNiu1999}, wavepacket propagation in
a slowly perturbed crystal has been described
using a combined Hamiltonian-Lagrangian approach. 
A derivation of the equations of motion using the Ehrenfest theorem          
without the Lagrangian formalism has been presented in
Ref.~\cite{ShindouImura2005}.
A series of articles by P.~Gosselin and coworkers has developed a purely
Hamiltonian semiclassical treatment~\cite{Gosselin2006, Gosselin2007,
Gosselin2008a}, and also perturbatively addressed the problem beyond the
semiclassical (lowest order in the reduced Planck constant $\hbar$)
approximation~\cite{Gosselin2008b, Gosselin2009}.
In other work, quantum kinetic equations have been derived for
multiband systems, taking the effects of phase-space Berry curvature
into account \cite{WongTserkovnyak2011}.

In this paper we approach the phase-space Berry curvature 
with applications in ultracold-atom systems in mind. 
We present two main results. 
First, we have derived quantum-mechanical 
Heisenberg equations of motion, where various Berry curvatures show up 
without relying on the semiclassical approximation,
Eqs.~\eqref{eqs:QuantumHeisenberg}.
These equations also allow us to recover the semiclassical results of
Ref.~\cite{SundaramNiu1999} by explicitly taking the small-curvature limit and
purely within the Hamiltonian formalism.
Second, we show that in the experimentally-accessible equal Rashba-Dresselhaus 
spin-orbit coupled system, the effective mass in the
semiclassical single-minimum regime can be reinterpreted as the 
phase-space Berry curvature, cf.~Fig.~\ref{fig1}.

This paper is organized as follows. 
In Sec.~\ref{sec:PosDepSOC} we present the problem and introduce the notation. 
In Sec.~\ref{sec:general}
we attack the general problem by performing an adiabatic approximation, 
which results in the Heisenberg equations of motion. 
To show that this treatment is also of practical interest,
we then apply these results to two particular cases in Sec.~\ref{sec:particular}. 
Namely, we derive the (quantum) equations of motion when either 
the position-space Berry curvature or the momentum-space Berry 
curvature is nonvanishing.
In Sec.~\ref{sec:semiclassical}, we perform the semiclassical approximation
for the general problem and obtain the corresponding equations of motion
with various Berry-curvature terms explicitly shown. 
We investigate the experimentally-relevant equal Rashba-Dresselhaus Hamiltonian
in our framework in Sec.~\ref{sec:spiral}.
Finally, Sec.~\ref{sec:summary} summarizes our results and provides
some directions for future work.

\section{Position-dependent spin-orbit coupling}
\label{sec:PosDepSOC}

Let us consider the following Hamiltonian with position-dependent spin-orbit coupling
\be
H
=
\frac{\hbar^{2}}{2m}( \bs k I - \bs A(\bs r) )^2
+
V(\bs r)
\,,
\label{eq:ham-1}
\ee
where $\bs A$ is a position-space vector of $2\times2$ matrices in the
spin space, $V$ is a spin-space matrix, and $I$ is the identity matrix. 
For concreteness and simplicity we consider (pseudo-)spin 1/2 atoms,
i.e., systems with two dispersion branches.  
Generalization to higher-spin systems with more than two dispersion
branches is straightforward and does not change the qualitative picture.
Note that $\bs A$ and $V$ may depend on position $\bs r$ \cite{Su2015}.
Wavevector $\bs k$ and
position $\bs r$ are (noncommuting) operators. We separate the spin-dependent
part of the potential,
\be
V(\bs r) =
\sum_{j} v^j(\bs r) \sigma^j+v_{0}({\bs r})I
\label{eq:scalar-s}
\ee
and also make the spin-dependence of the vector potential $\bs A$ explicit,
\be
A_j(\bs r)=\sum_{l} a_j^l({\bs r}) \sigma^l\,,
\label{eq:vector-s}
\ee
where $\sigma^j$ are the Pauli matrices. Note that the
square of the $\bs A$ matrix is proportional to the identity matrix.
We write position-space vectors using bold font and their 
indices as subscripts, whereas spin-space vectors are denoted by
the arrow above and their indices are written as superscripts. 
The matrix $\bs A$ may also contain a term proportional to the identity
matrix.  Such a term would describe the usual $U(1)$ electromagnetic field,
which is beyond the scope of this paper, and we thus neglect it.

Using the matrices given in Eqs.~(\ref{eq:scalar-s})
and (\ref{eq:vector-s}), the Hamiltonian in Eq.~(\ref{eq:ham-1}) becomes
\ba
H=
&
\frac{\hbar^{2}}{2m}
I
\sum_i k_i k_i
+
\sum_{j}
\Big(
-\frac{\hbar^{2}}{2m}
\sum_i \{k_i,a_i^j\} + v^j
\Big)
\sigma^j
\nonumber \\ &
+
\Big(
\sum_{i,j} \frac{\hbar^{2}}{2m} a_i^j a_i^j 
+v_{0}
\Big)
I\,,
\label{eq:ham-2}
\ea
where we have introduced the anticommutator
$\{k_i,a_i^j\}=k_i a_i^j+a_i^j k_i$.
The Hamiltonian above
contains a term of the Zeeman form, and thus it is natural to
introduce the operator
\be
B^j = -\frac{\hbar^{2}}{2m}
\sum_i \{k_i,a_i^j\} + v^j \,,
\ee
which plays the role of the magnetic field in this term. 
We will therefore use the term ``effective magnetic field'' to
describe the operator $B^j$ from here on.
Furthermore, there is a spin-independent potential in the Hamiltonian,
\be
W(\bs r)
=
\frac{\hbar^{2}}{2m}
\sum_{i,j} [a_i^j(\bs r)]^2
+
v_0(\bs r)\,.
\ee
We now are in the position to write down 
the initial Hamiltonian in the following
concise manner,
\begin{equation}
 H=
\frac{\hbar^{2}}{2m}{\bs k}^{2} I
+\BdotSigma
+W(\bs r) I
\end{equation}
We proceed to look for the solution of the time-dependent 
Schr\"odinger equation 
\be
i\hbar\frac{\partial}{\partial t}\Psi
=
 H\Psi
\label{eq:schroed}
\ee
using adiabatic approximation by assuming that the distance between
the eigenvalues of the operator 
$\BdotSigma$
is large compared to the off-diagonal terms.

\section{General equations for adiabatic approximation}
\label{sec:general}

In this Section we treat the quantum mechanical problem exactly first,
and then in adiabatic approximation. We do not
directly perform, for example, the expansion in orders of $\hbar$
as carried out in Refs.~\cite{Gosselin2006, Gosselin2007, 
Gosselin2008a, Gosselin2008b, Gosselin2009}. 
Instead, we postpone 
the semiclassical approximation to Sec.~\ref{sec:semiclassical} 
of the paper.

\subsection{Unitary transformation}

Anticipating adiabatic approximation, 
let us define a unitary operator $U$, which diagonalizes the term
$\BdotSigma$
in spin space. 
Our problem is divided up into two dispersion branches which we label 
with the sign of the eigenvalue of the operator $\vec{B}\cdot\vec{\sigma}$.
In particular, the definition of $U$ implies that
\be
\mathcal{P}_{+}^{\dag}{U}^{\dag}\BdotSigma U\mathcal{P}_{-}
=
\mathcal{P}_{-}^{\dag}{U}^{\dag}\BdotSigma U\mathcal{P}_{+}
=0\,,
\ee
where
\be
\mathcal{P}_{+}=\left(\begin{array}{c}
1\\
0
\end{array}\right)\,,\qquad\mathcal{P}_{-}=\left(\begin{array}{c}
0\\
1
\end{array}\right)\,,
\ee
are the respective $\sigma^z$ eigenstates.
Therefore, we can rewrite the diagonalized Zeeman term as
\be
{U}^{\dag} \BdotSigma U
=
\mathcal{P}_{+}^{\dag}{U}^{\dag}\BdotSigma{U}\mathcal{P}_{+}+
\mathcal{P}_{-}^{\dag}{U}^{\dag}\BdotSigma{U}\mathcal{P}_{-}
\,.
\ee
The wavefunction in the diagonal basis is related to the original
wavefunction by the same transformation,
\be
\tilde{\Psi}={U}^{\dag}\Psi \,. 
\ee
Plugging this definition into Eq.~\eqref{eq:schroed} yields
the Schr\"odinger equation in the new basis,
\be
i\hbar\frac{\partial}{\partial t}\tilde{\Psi}
=
\tilde{H}\tilde{\Psi}\,,
\label{eq:schroed-2}
\ee
where from
\be
\tilde{H}={U}^{\dag}H{U}
\ee
we see that the Hamiltonian after the transformation
retains its original form,
\be
\tilde{H}=
\frac{\hbar^{2}}{2m}\tilde{{\bs k}}^{2} I
+\TildeBdotSigma
+W(\tilde{{\bs r}})I
\ee
and the effect of this transformation can be incorporated
through a redefinition of the effective magnetic field, position, momentum,
and spin operators, namely,
\bsa
\tilde {\bs r} &= U^\dag \bs r U\,,\\
\tilde {\bs k} &= U^\dag \bs k U\,,\\
\vec {\tilde \sigma} &= U^\dag \vec \sigma U\,.
\esa
We note that though $\TildeBdotSigma$ is proportional to $\sigma_z$,
in general both $\vec{\tilde B}$ and $\vec {\tilde \sigma}$ have components
in all three directions, and only their scalar product is diagonal
in spin space. 

\subsection{Adiabatic approximation}

Thus far our discussion has been exact. 
Let us now perform adiabatic approximation by assuming 
that the wavefunction $\tilde{\Psi}$ remains in the
eigenspace of the projection operator 
$\mathcal{P}_{\pm}\mathcal{P}_{\pm}^{\dag}$, i.e., either in 
the lower or the upper dispersion branch with respect to the position- and
momentum-dependent effective magnetic field.
Explicitly, $\tilde{\Psi}=\psi\mathcal{P}_{\pm}$ defines
another wavefunction $\psi$, the components of which now evolve according
to the Schr\"odinger equation with an effective Hamiltonian
\be
H_{\mathrm{eff}}=\mathcal{P}_{\pm}^{\dag}\tilde{H}\mathcal{P}_{\pm}
\ee
either in the lower ($-$) or the upper ($+$) branch. 
In the effective Hamiltonian 
the operators 
\bsa
{\bs r}_{\mathrm{c}} & 
=\mathcal{P}_{\pm}^{\dag}\tilde{\bs r}\mathcal{P}_{\pm}
=\mathcal{P}_{\pm}^{\dag}{U}^{\dag}{\bs r}{U}\mathcal{P}_{\pm}\,,\\
{\bs k}_{\mathrm{c}} & 
=\mathcal{P}_{\pm}^{\dag}\tilde{\bs k}\mathcal{P}_{\pm}
=\mathcal{P}_{\pm}^{\dag}{U}^{\dag}{{\bs k}}{U}\mathcal{P}_{\pm}\,,
\esa
appear. They describe the position and momentum operators adiabatically
projected to one of the branches. 
These operators $\bs r_\mathrm{c}$ and $\bs k_\mathrm{c}$ are sometimes
called the covariant operators \cite{Gosselin2006}. They
are manifestly different from their canonical counterparts, signalling
breakdown of the Galilean invariance \cite{ZhangEtAl2016}.
Note that even though $\bs r_\mathrm{c}$ can be understood as
a physical position operator (i.e., the position operator describing, e.g., 
the motion of the center of a wavepacket), $\bs k_\mathrm{c}$ does
not correspond to kinetic momentum. A kinetic momentum,
also known as physical momentum \cite{ZhangEtAl2016}, operator
could be obtained by performing the transformations described here
on $\hbar (\bs k I - \bs A)$, which is a matrix in spin space, as opposed to 
merely $\hbar \bs k$. One can convince oneself that $\hbar(\bs k I - \bs A)$ 
is the kinetic momentum by computing the commutator between
$\bs r_\mathrm{c}$ and the Hamiltonian in Eq.~\eqref{eq:ham-1}.

Since the position operator $\bs r$ and the momentum operator
$\hbar\bs k$ are both spin independent, the projection operator commutes
with them. Using this property, the last two equations can be rewritten as
\bsa
\label{eq:rc}
\bs r_{\mathrm{c}} & =\bs r - \bs{\mathcal{A}}^{(k)}\,,\\
\bs k_{\mathrm{c}} & =\bs k - \bs{\mathcal{A}}^{(r)}\,,
\label{eq:kc}
\esa
where 
\begin{subequations}
\label{eqs:AkAr}
\ba
\bs{\mathcal{A}}^{(k)} 
& =
-\mathcal{P}_{\pm}^{\dag}{U}^{\dag}[{\bs r},{U}]\mathcal{P}_{\pm}\,, 
\\
\bs{\mathcal{A}}^{(r)} 
& =
-\mathcal{P}_{\pm}^{\dag}{U}^{\dag}[{{\bs k}},{U}]\mathcal{P}_{\pm}\,.
\ea
\end{subequations}
Operators $\bs{\mathcal{A}}^{(k)}$ and $\bs{\mathcal{A}}^{(r)}$ represent the
Berry connections. 
Given a suitable representation, commutators in these operators become
derivatives, e.g., a commutator of a function with the position operator
is proportional to a momentum derivative in the momentum representation. 
Therefore, when the operator $U$ is diagonal in position or momentum basis,
the Berry connection operator $\bs{\mathcal{A}}^{(k)}$ or
$\bs{\mathcal{A}}^{(r)}$ becomes a connection in the usual geometric sense,
see Eqs.~\eqref{eq:Ar-position} and \eqref{eq:Ak-momentum} below. 
This also explains the seemingly counter-intuitive labeling, which is
standard \cite{XiaoEtAl2010}.

In order to evaluate the potential term
it is beneficial to expand $W({\bs r})$ in a power series,
\be
W({\bs r})=w^{(0)}+\sum_{j}w_{j}^{(1)}r_{j}+\sum_{j,l}w_{jl}^{(2)}r_{j}r_{l}+\cdots\,.
\ee
In most ultracold-atom-related problems the potential $W$ is at most
quadratic, and we will therefore limit our attention to such cases. 
Cubic and higher order terms in the potential $W$ would result in
a more complicated expression in Eq.~\eqref{eq:Vk}.
The effective Hamiltonian $H_{\mathrm{eff}}$ can thus be rewritten
in the following concise form:
\begin{equation}
H_{\mathrm{eff}}=\frac{\hbar^{2}}{2m} {\bs k}_\mathrm{c}^2
+
W({\bs r}_\mathrm{c})
+\mathcal{V}\,,
\end{equation}
where
\be
\label{eq:PotDef}
\mathcal{V}
=\mathcal{P}_{\pm}^{\dag}
\TildeBdotSigma
\mathcal{P}_{\pm}+\mathcal{V}^{(r)}+\mathcal{V}^{(k)}\,,
\ee
with
\begin{subequations}
\label{eqs:VrVk}
\ba
\mathcal{V}^{(r)} & =\frac{\hbar^{2}}{2m}
\big(
\mathcal{P}_{\pm}^{\dag}{U}^{\dag}[{{\bs k}},{U}]\mathcal{P}_{\mp}
\big)
\cdot
\big(
\mathcal{P}_{\mp}^{\dag}{U}^{\dag}[{{\bs k}},{U}]\mathcal{P}_{\pm}
\big)
\,,\\
\label{eq:Vk}
\mathcal{V}^{(k)} & =\sum_{j,l}w_{jl}^{(2)}\mathcal{P}_{\pm}^{\dag}{U}^{\dag}[r_{j},{U}]\mathcal{P}_{\mp}\mathcal{P}_{\mp}^{\dag}{U}^{\dag}[r_{l},{U}]\mathcal{P}_{\pm}\,.
\ea
\end{subequations}
We see that besides the alternation of the physical momentum and position
operators, three extra potential terms have appeared. 
In the following subsections we investigate dynamics in this system
in more detail.

\subsection{Heisenberg equations}

As discussed above, the operators ${\bs r}_{\mathrm{c}}$ and 
$\hbar{\bs k}_{\mathrm{c}}$ do not represent the canonical 
position and momentum. This is confirmed by the observation that
their commutators differ from the usual commutators for the 
position and momentum operators.
In particular, not only the position-momentum commutator gains an extra
term, but the other commutators do not vanish anymore:
\begin{subequations}
\label{eqs:Comms}
\ba
[(r_{\mathrm{c}})_{j},(r_{\mathrm{c}})_{l}] & 
=i\Theta_{jl}^{(k,k)}\,,\\
[(k_{\mathrm{c}})_{j},(k_{\mathrm{c}})_{l}] & 
=i\Theta_{jl}^{(r,r)}\,,\\
[(r_{\mathrm{c}})_{j},(k_{\mathrm{c}})_{l}] &
=i\delta_{j,l}+i\Theta_{jl}^{(k,r)}\,,
%
\ea
\end{subequations}
where various Berry curvatures are given by 
\begin{subequations}
\label{eqs:BerryCurvatures}
\ba
\Theta_{jl}^{(k,k)} &
=i[r_{j},\mathcal{A}_{l}^{(k)}]-i[r_{l},\mathcal{A}_{j}^{(k)}]\,,\\
\Theta_{jl}^{(r,r)} &
=i[{k}_{j},\mathcal{A}_{l}^{(r)}]-i[{k}_{l},\mathcal{A}_{j}^{(r)}]\,,\\
\Theta_{jl}^{(k,r)} &
=i[r_{j},\mathcal{A}_{l}^{(r)}]-i[{k}_{l},\mathcal{A}_{j}^{(k)}]\,,\\
\Theta_{jl}^{(r,k)} &
=i[{k}_{j},\mathcal{A}_{l}^{(k)}]-i[r_{l},\mathcal{A}_{j}^{(r)}]\,,
\ea
\end{subequations}
and also note that $\Theta_{jl}^{(k,r)} = - \Theta_{lj}^{(r,k)}$.
The emergence of the extra terms in Eqs.~\eqref{eqs:Comms} 
can be interpreted as curving up of 
position space, momentum space, and phase space, respectively.
In quantum mechanics phase space is inherently curved,
as position and momentum operators do not commute to begin with.
Eqs.~\eqref{eqs:Comms} demonstrate that non-commutative geometry 
underlies the algebraic structure of coordinates and momenta. 
Non-commutative coordinates in the context of field theory, and physics
in general, have attracted a great deal of theoretical interest
\cite{SeibergWitten1999, DouglasNekrasov2001}.

Adiabatic approximation, and spin projection
in particular, is essential to obtaining a nonzero Berry curvature
(see Ref.~\cite{FujitaEtAl2011} for an extensive discussion).
If no spin projection is performed, Berry connections generally have a
nontrivial matrix structure. Schematically, this matrix structure serves to 
generate terms of the $[\mathcal A, \mathcal A]$ type, which exactly
cancel the corresponding $[k,\mathcal A]$ and $[r,\mathcal A]$ terms.
In this way it is ensured that a change of basis in spin space leaves
physical dynamics unchanged.
On the other hand, adiabatic approximation confines evolution 
of the system to only one of the two dispersion branches, and the price for 
that simplification is the emergence of Berry curvature.

Going beyond adiabatic approximation would bring about
effects similar to Zitterbewegung, see 
Refs.~\cite{Merkl08, LeBlanc2013,WinklerEtAl2007}.
In that case, noncommuting components of the velocity operator
$\dot {\bs r}_\mathrm{c}$ signal
interbranch transitions. The frequency of these transitions is
given by the gap between the dispersion branches.

Naturally, modified commutation relations result in altered
Heisenberg equations for the covariant operators
\bsa
\dot {\bs k}_{\mathrm{c}} & 
=  \frac{1}{i\hbar}[{\bs k}_{\mathrm{c}},H_{\mathrm{eff}}]\,,\\
\dot {\bs r}_{\mathrm{c}} & 
=  \frac{1}{i\hbar}[{\bs r}_{\mathrm{c}},H_{\mathrm{eff}}]\,,
\esa
which now contain the Berry curvature terms defined above,
\begin{subequations}
\label{eqs:QuantumHeisenberg}
\ba
\dot {\bs k}_{\mathrm{c}}  = & 
-\frac{1}{\hbar}\bs{\nabla}W({\bs r}_{\mathrm{c}})
+\frac{1}{i\hbar}[{\bs k}_{\mathrm{c}},\mathcal{V}]
\\
& + \frac{\hbar}{2m}\sum_{j,l} \bs e_j
\{ \Theta_{jl}^{(r,r)} , (k_{\mathrm{c}})_{l} \}
\nonumber \\
&+\frac{1}{2\hbar}\sum_{j,l}
\bs e_j
\{ \Theta_{jl}^{(r,k)}, \nabla_l W(\bs r_\mathrm{c}) \}
\,, \nonumber \\
\dot {\bs r}_{\mathrm{c}}  = &
\frac{\hbar}{m}{\bs k}_{\mathrm{c}}
+\frac{1}{i\hbar}[{\bs r}_{\mathrm{c}},\mathcal{V}]
\\
& +\frac{\hbar}{2m}\sum_{j,l}
\bs e_j
\{\Theta_{jl}^{(k,r)}, (k_{\mathrm{c}})_{l} \}
\nonumber \\
&+\frac{1}{2\hbar}\sum_{j,l}
\bs e_j
\{ \Theta_{jl}^{(k,k)}, \nabla_l W (\bs r_\mathrm{c}) \}
\,, \nonumber
\ea
\end{subequations}
where $\bs e_j$ denotes the unit vector in the $j$ direction (i.e.,
$\bs e_j$ is a vector, and not a component of a vector), while
all the $\nabla$'s here act on functions of a single variable, and
thus represent derivatives with respect to that variable.
This set of Eqs.~\eqref{eqs:QuantumHeisenberg} 
represents the first main result of the paper.

The first term on the right-hand side in each of the two equations denotes the
usual contribution known from classical mechanics, whereas the subsequent terms
are due to adiabatic approximation. In particular, in each case the second
term is due to emergent potentials, whereas the last two terms are due to Berry
curvatures.
In order to make our hitherto abstract discussion more concrete,
we consider two simple examples next.

\section{Particular cases}
\label{sec:particular}

Let us consider the situation where the operator $\vec B$ 
becomes a vector of complex numbers (as opposed to operators) in some
representation.
In
this case the eigenvalues of the operator $\BdotSigma$
are $\pm|\vec B|$ with the eigenfunctions
\be
\chi_{+}=\left(\begin{array}{c}
e^{-i\frac{\phi}{2}}\cos\frac{\alpha}{2}\\
e^{i\frac{\phi}{2}}\sin\frac{\alpha}{2}
\end{array}\right)\,,\quad\chi_{-}=\left(\begin{array}{c}
e^{-i\frac{\phi}{2}}\sin\frac{\alpha}{2}\\
-e^{i\frac{\phi}{2}}\cos\frac{\alpha}{2}
\end{array}\right)\,,\label{eq:spinor}
\ee
where the spherical angles $\alpha$ and $\phi$ give the direction
of the vector $\vec B$. The unitary operator ${U}$, which diagonalizes
the matrix $\BdotSigma$, then consists 
of two columns, $\chi_{+}$ and $\chi_{-}$. 
Two particular cases, which often occur in practice, are a position-dependent
effective magnetic field, and spin-orbit coupling which corresponds to a
momentum-dependent Zeeman term. We investigate them in more detail below.

\subsection{Position space curvature}

When the operator $\vec B$ depends only on the coordinate 
$\bs r$, the unitary transformation operator is a function of position only:
$U = U(\bs r)$. 
In this case the
Berry connections are $\bs{\mathcal{A}}^{(k)}=0$ and 
\be
\bs {\mathcal{A}}^{(r)}
=i\chi_{\pm}^{\dag}(\bs r)
\bs {\nabla}^{(r)}
\chi_{\pm}(\bs r)
\,.
\label{eq:Ar-position}
\ee
The corresponding Berry curvatures are $\Theta^{(k,k)}=\Theta^{(k,r)}=\Theta^{(r,k)}=0$
and
\be
\Theta_{jl}^{(r,r)}=
\nabla_j^{(r)} \mathcal{A}_{l}^{(r)}
-\nabla_l^{(r)} \mathcal{A}_{j}^{(r)}
\,.
\ee
The scalar potentials are $\mathcal{V}^{(k)}=0$ and
\begin{equation}
\mathcal{V}^{(r)}
=-\frac{\hbar^{2}}{2m}
\left(
\chi_{\pm}^{\dag}\bs \nabla^{(r)}\chi_{\mp}
\right)
\cdot
\left(
\chi_{\mp}^{\dag}\bs{\nabla}^{(r)}\chi_{\pm}
\right)
\,.
\end{equation}
These quantities can also be expressed in terms of the
angles on the Bloch sphere, namely, $\alpha$ and $\phi$.
Concretely, the Berry connection is
\be
\bs {\mathcal A}^{(r)} = \pm \frac{1}{2} \cos\alpha \bs \nabla^{(r)}\phi\,,
\ee
the potential is the same for the two branches,
\be
\mathcal V^{(r)} = 
\frac{\hbar^2}{8m}
\Big(
(\bs \nabla^{(r)} \alpha)^2
+
\sin^2 \alpha (\bs \nabla^{(r)} \phi)^2
\Big)\,,
\label{eq:PotentialUsingAngles}
\ee
while the Berry curvature is opposite for the two brances,
\ba
\Theta^{(r,r)}_{jl} = &
\pm \frac{\sin \alpha}{2}
\Big(
(\nabla^{(r)}_j \phi)
(\nabla^{(r)}_l \alpha)
\nonumber \\ &
-
(\nabla^{(r)}_j \alpha)
(\nabla^{(r)}_l \phi)
\Big)
\,.
\ea
The latter two equations can also be conveniently expressed
using the unit vector $\vec n = \vec B / |\vec B|$,
describing the direction of the vector $\vec B$ \cite{Volovik1987},
\ba
\mathcal{V}^{(r)} & = \frac{\hbar^{2}}{8m}
(\bs{\nabla}^{(r)}\vec{n})^{2}
\,, \\
\Theta^{(r,r)}_{jl} & =
\mp\frac{1}{2}
\vec{n}
\cdot
(\nabla_j^{(r)}\vec{n}
\times
\nabla_l^{(r)}\vec{n})
\,.
\label{eq:CurvatureUsingUnitVector}
\ea
The equations of motion \eqref{eqs:QuantumHeisenberg} in this case are
\bsa
\dot {\bs k}_{\mathrm{c}}  = & 
-\frac{1}{\hbar}\bs\nabla^{(r)}W({\bs r}_{\mathrm{c}})
+\frac{1}{i\hbar}[{\bs k}_{\mathrm{c}},\mathcal{V}]
\\
& + \frac{\hbar}{2m}\sum_{j,l} \bs e_j
\{ \Theta_{jl}^{(r,r)} , (k_{\mathrm{c}})_{l} \}
\,, \nonumber \\
\dot {\bs r}_{\mathrm{c}}  = &
\frac{\hbar}{m}{\bs k}_{\mathrm{c}}
\,.
\esa
The curvature $\Theta_{jl}^{(r,r)}$ is also known as the synthetic magnetic
field and has been studied theoretically
\cite{DalibardEtAl2011,GoldmanEtAl2014} as well as experimentally
\cite{LinEtAl2009a,BeelerEtAl2013}.
Note that the Stern-Gerlach experiment \cite{GerlachStern1922}, which
is often discussed in introductory quantum mechanics textbooks
\cite{Sakurai1994}, can also be described in this framework. In particular, 
a linear magnetic field gradient separates out the two spin components
since the Berry connection in position space becomes nonzero, even though
the Berry curvature itself vanishes in that case. 

\subsection{Momentum space curvature}

When operator $\vec B$ depends only on momentum $\bs k$,
basis transformation is facilitated by $U=U(\bs k)$, which
then is a function of momentum only. In
this case, the position-space Berry connection vanishes, 
$\bs {\mathcal{A}}^{(r)}=0$, whereas
\begin{equation}
\boldsymbol{\mathcal{A}}^{(k)}
=
-i
\chi_\pm^{\dag}
\bs{\nabla}^{(k)}
\chi_\pm
\,.
\label{eq:Ak-momentum}
\end{equation}
Accordingly, the Berry curvature components related to 
position space vanish, $\Theta^{(r,r)}=\Theta^{(k,r)}=\Theta^{(r,k)}=0$,
and all the curvature is concentrated in the momentum space,
\be
\Theta_{jl}^{(k,k)}
=
-\nabla_j^{(k)}\mathcal{A}_{l}^{(k)}
+\nabla_l^{(k)}\mathcal{A}_{j}^{(k)}
\,.
\ee
The scalar potential has momentum components only, i.e.,
$\mathcal{V}^{(r)}=0$ and
\be
\mathcal{V}^{(k)}
=-\sum_{j,l}w_{jl}^{(2)}
\left(
\chi_{\pm}^{\dag} \nabla^{(k)}_j\chi_{\mp}
\right)
\left(
\chi_{\mp}^{\dag} \nabla^{(k)}_l \chi_{\pm}
\right)
\,.
\ee
Equations similar to 
Eqs.~\eqref{eq:PotentialUsingAngles} -- \eqref{eq:CurvatureUsingUnitVector}
can also be written down in the case at hand.
Finally, Heisenberg equations of motion \eqref{eqs:QuantumHeisenberg} 
in this case are
\bsa
\dot {\bs k}_{\mathrm{c}}  = & 
-\frac{1}{\hbar} \bs \nabla^{(r)} W({\bs r}_{\mathrm{c}})
\,, \\
\dot {\bs r}_{\mathrm{c}}  = &
\frac{\hbar}{m}{\bs k}_{\mathrm{c}}
+\frac{1}{i\hbar}[{\bs r}_{\mathrm{c}},\mathcal{V}]
\\
&+\frac{1}{2\hbar}\sum_{j,l}
\bs e_j
\{ \Theta_{jl}^{(k,k)}, \nabla^{(r)}_l W (\bs r_\mathrm{c}) \}
\,. \nonumber
\esa

We remark that strictly speaking 
when the two dispersion branches
touch, adiabatic approximation becomes invalid. The condition for the validity 
of adiabatic approximation is particularly elegant in this
momentum-space curvature case. Concretely, $|\vec B|$
must be nonzero for each $\bs k$. This only is the case when the
system of equations
\be
-\frac{\hbar^2}{m} \sum_i k_i a_i^j + v^j = 0
\ee
has no solutions. In terms of $\bs k$, this system of equations
is linear. Therefore, the branches do not touch only if
the following determinant in spin space vanishes:
\be
\sum_{j,l,n}\epsilon^{jln} a_1^j a_2^l a_3^n = 0,
\ee
where $\epsilon^{jln}$ is the Levi-Civita symbol.
Effects of the momentum-space curvature have already been extensively
studied theoretically \cite{PriceEtAl2014,PriceEtAl2015b}.

\section{Semiclassical approximation}
\label{sec:semiclassical}

As we have seen in the last section, in cases where all the Berry curvature 
is concentrated in a single component, for example, as either 
position-space curvature or momentum-space curvature, the equations 
of motion are relatively simple and can often be treated exactly.
However, this is not the case in general. 
Moreover, in the definition of the Berry connection or the 
Berry curvatures, $\hbar$ is crucially absent, suggesting that 
these quantities also affect semiclassical dynamics.
Finally, semiclassical approximation is of practical interest,
as it allows one to investigate the motion of wavepackets
and clouds of ultracold atoms in particular.
Motivated by these points, we consider semiclassical approximation
in general in this Section, and proceed to apply it to particular
Hamiltonians subsequently.

In semiclassical approximation we neglect the commutator between 
position and momentum. In that case the effective magnetic field 
operator becomes 
\be
B^j = -\frac{\hbar^{2}}{m}
\sum_i k_i a_i^j(\bs r) + v^j(\bs r) \, .
\ee
The matrix $\BdotSigma$ still
has the eigenvectors $\chi_\pm$, but they now parametrically
depend on the numbers $\bs r$ and $\bs k$. Berry
connections \eqref{eqs:AkAr} are
\bsa
\bs {\mathcal A}^{(r)} 
&= i \chi_\pm^\dagger \bs \nabla^{(r)} \chi_\pm \,,
\\
\bs {\mathcal A}^{(k)} 
&= -i \chi_\pm^\dagger \bs \nabla^{(k)} \chi_\pm \,,
\esa
and lead to corresponding Berry curvatures \eqref{eqs:BerryCurvatures}
\bsa
\Theta_{jl}^{(k,k)} & =-\nabla^{(k)}_j \mathcal{A}_{l}^{(k)}
+\nabla^{(k)}_l \mathcal{A}_{j}^{(k)}\,,\\
\Theta_{jl}^{(r,r)} & =\nabla^{(r)}_j \mathcal{A}_{l}^{(r)}
-\nabla^{(r)}_l \mathcal{A}_{j}^{(r)}\,,\\
\Theta_{jl}^{(k,r)} & =-\nabla^{(k)}_j \mathcal{A}_{l}^{(k)}
-\nabla^{(r)}_l \mathcal{A}_{j}^{(r)}\,,\\
\Theta_{jl}^{(r,k)} & =\nabla^{(r)}_j \mathcal{A}_{l}^{(r)}
+\nabla^{(k)}_l \mathcal{A}_{j}^{(k)}\,,
\esa
and scalar potentials \eqref{eqs:VrVk} 
\bsa
\mathcal{V}^{(r)} & =-\frac{\hbar^{2}}{2m}
\left(
\chi_{\pm}^{\dag} \bs \nabla^{(r)} \chi_{\mp}
\right)
\cdot
\left(
\chi_{\mp}^{\dag} \bs \nabla^{(r)} \chi_{\pm}
\right)
\,,\\
\mathcal{V}^{(k)} & =
-\sum_{j,l}w_{jl}^{(2)}
\chi_{\pm}^{\dag} \nabla^{(k)}_j \chi_{\mp}
\chi_{\mp}^{\dag} \nabla^{(k)}_l \chi_{\pm}
\,.
\esa
Equations of motion \eqref{eqs:QuantumHeisenberg} then take the form
\begin{subequations}
\label{eqs:SemiClassical}
\ba
\dot {\bs k}_{\mathrm{c}}  = & 
-\frac{1}{\hbar}\bs \nabla^{(r)}W 
-\frac{1}{\hbar}\bs \nabla^{(r)} \mathcal{V}
\\
& + \frac{\hbar}{m}\sum_{j,l} \bs e_j
 \Theta_{jl}^{(r,r)} (k_{\mathrm{c}})_{l}
\nonumber \\
&+\frac{1}{\hbar}\sum_{j,l}
\bs e_j
\Theta_{jl}^{(r,k)} \nabla^{(r)}_l W
\,, \nonumber \\
\dot {\bs r}_{\mathrm{c}}  = &
\frac{\hbar}{m}{\bs k}_{\mathrm{c}}
+\frac{1}{\hbar} \bs \nabla^{(k)} \mathcal{V}
\\
& +\frac{\hbar}{m}\sum_{j,l}
\bs e_j
\Theta_{jl}^{(k,r)} (k_{\mathrm{c}})_{l}
\nonumber \\
&+\frac{1}{\hbar}\sum_{j,l}
\bs e_j
\Theta_{jl}^{(k,k)} \nabla^{(r)}_l W
\,. \nonumber
\ea
\end{subequations}
%
We see that in semiclassical approximation Berry curvatures $\Theta$
and scalar potentials $\mathcal{V}$ do not vanish and still show up in the
equations of motion. However, since they involve derivatives of the
eigenvectors $\chi_\pm$, which
vary at scales much larger than the other relevant length scales, e.g., 
the de~Broglie wavelength, these quantities are not dominant. 
Therefore, it is a reasonable approximation to insert classical relations
$\dot{\bs k}_{\mathrm{c}} \approx - \bs \nabla^{(r)} W/\hbar$
and
$\dot{\bs r}_{\mathrm{c}} \approx {\hbar}{\bs k}_{\mathrm{c}}/m$
on the right-hand side of the equations of motion. In that case, we arrive at
\begin{subequations}
\label{eqs:SemiClassicalSmallCurv}
\ba
\dot{\bs k}_{\mathrm{c}}  = & 
-\frac{1}{\hbar}\bs \nabla^{(r)}( W + \mathcal V )
\\
& 
+ \sum_{j,l} \bs e_j
\left(
 \Theta_{jl}^{(r,r)} (\dot r_{\mathrm{c}})_{l}
+\Theta_{lj}^{(k,r)} (\dot k_{\mathrm{c}})_{l}
\right)
\,, \nonumber \\
\dot {\bs r}_{\mathrm{c}}  = &
\frac{\hbar}{m}{\bs k}_{\mathrm{c}}
+\frac{1}{\hbar} \bs \nabla^{(k)} \mathcal{V}
\\
& -\sum_{j,l}
\bs e_j
\left(
\Theta_{lj}^{(r,k)} (\dot r_{\mathrm{c}})_{l}
+\Theta_{jl}^{(k,k)} (\dot k_{\mathrm{c}})_{l}
\right)
\,, \nonumber
\ea
\end{subequations}
where we have grouped the terms in order to facilitate comparison with
Eqs.~(2.19) in Ref.~\cite{SundaramNiu1999}.  
Even though the result is the
same, we arrived at it purely within the Hamiltonian formalism.
Furthermore, in our derivation we have shown that these equations constitute a
special case of the semiclassical situation, namely, they are only valid when
Berry curvatures are small.

\section{One-dimensional spiral}
\label{sec:spiral}

Up to this point our discussion was valid for an arbitrary number of spatial
dimensions. Now we specialize to a single spatial dimension, where the effects
of phase space Berry curvature are nevertheless nontrivial, as phase space in
this case is two dimensional.  In particular, as an application of the
discussion above, in this Section we investigate the dynamics described by the
following one-dimensional Hamiltonian:
\be
H = H_0 - FxI\,,
\ee
where the $-FxI$ term describes a spin-independent force
(linear potential), while
\ba
H_0  = 
&
\frac{\hbar^{2}}{2m}
(kI-a\sigma_{3})^2
\label{eq:h0}
\\ \nonumber
&
+
\frac{\hbar\rabi}{2}
\Big(
\cos\frac{x}{\lambda}\, \sigma_1 + \sin\frac{x}{\lambda}\, \sigma_{2}
\Big)
\ea
is an exactly-solvable Hamiltonian with a constant spin-orbit coupling term
and a position-dependent Zeeman term. The spin-orbit coupling term
is characterized by its constant strength $a$. The position-space Zeeman
term constitutes a spiral with a constant magnitude $\rabi / 2$ and 
a wavelength $\lambda$. Furthermore, it is convenient to introduce the
wavenumber
\be
\kappa=
\frac{\rabi}{2a}
\frac{m}{\hbar}\,,
\ee
which quantifies the relative strength of the position-space
and momentum-space Zeeman terms.
This Hamiltonian $H_0$ is well known: it has undergone 
an extensive theoretical investigation (see Ref.~\cite{LiEtAl2015} and references
therein) and has also been realized experimentally \cite{LinEtAl2011a}.

\subsection{Exact solution}

In this subsection we recap the exact diagonalization 
\cite{LiEtAl2015} of $H_0$. We
begin by observing that the Zeeman term can be rewritten as 
\be
\cos(x/\lambda)\sigma_{1}
+\sin(x/\lambda)\sigma_{2}
=
e^{-i\frac{x}{2\lambda}\sigma_{3}}\sigma_{1}e^{i\frac{x}{2\lambda}\sigma_{3}}\,,
\ee
and hence
the coordinate dependence may be eliminated by performing a unitary
transformation. Concretely, the wave function in the new basis is 
\be
\tilde{\Psi}=U^{\dag}\Psi \,,
\ee
where
\be
U=e^{-i\frac{x}{2\lambda}\sigma_{3}} \,.
\ee
We then obtain the transformed Hamiltonian
\be
\tilde{H}_0=U^{\dag}H_0U=
\frac{\hbar^{2}}{2m}
(k-k_0\sigma_{3})^{2}
+\frac{\hbar\rabi}{2}\sigma_{1}
\label{eq:h0tilde}
\,,
\ee
where 
\be
k_0=a+\frac{1}{2\lambda}
\label{eq:k0}
\,.
\ee
The symbols in this Hamiltonian directly correspond to
physical quantities in the experimental realization of the model
\cite{LinEtAl2011a,LiEtAl2015}.
In particular, given two lasers with small detuning from the Raman
resonance, $k_0$ is the wavevector difference of the lasers
(recoil wavevector), and $E_R = \hbar^2 k_0^2/2m$ is the recoil energy.
The intensity of the two lasers is represented by the Rabi frequency
$\rabi$.
The dispersion is thus given by
\be
\frac{E}{\hbar^2/m}= \frac{k^2}{2} + \frac{k_0^2}{2}
\pm
\sqrt{ (k k_0)^2 + \frac{\rabi^2}{4}\left(\frac{m}{\hbar}\right)^2 }
\ee
in the lower ($-$) and the upper ($+$) branch.
The upper branch has a single quadratic minimum, 
wheras the lower branch has one quadratic minimum when
\be
\hbar \rabi > 4 E_R
\label{eq:oneMinimum}
\,,
\ee
and two quadratic minima or one quartic minimum otherwise.
As the gap between the two branches is governed by the
ratio $\hbar\rabi/E_R$,
and we will be interested in the situation where the
adiabatic approximation is applicable, we limit further discussion
to the single minimum regime given in Eq.~\eqref{eq:oneMinimum}. 
Expanding the dispersion
around the minimum $k=0$, we find that the effective mass $m^*_\pm$ 
in the two branches is given by
\be
\frac{m}{m^*_\pm} 
= 1\pm\frac{2 k_0^2}{\rabi} \frac{\hbar}{m}
\approx 1\pm\frac{1}{\kappa}\left(a+\frac{1}{\lambda}\right)
\label{eq:EffectiveMass}
\,,
\ee
respectively. The first equality applies in the limit $k \ll k_0$
of interest when computing the effective mass, whereas the
second approximation holds in the semiclassical regime.
Indeed, the smallness of $1/\lambda a$ and $1/\lambda\kappa$ 
defines the semiclassical regime, where all the gradients are small
compared to the de~Broglie wavelength. Namely,
when the period of the spiral $\lambda$ in the position-space Zeeman term 
is large as compared to the spin-orbit coupling wavenumber $a$,
we have $1/\lambda a \ll 1$. In combination to the single-minimum
condition, this also implies $1/\lambda\kappa \ll 1$.
Note further that one can transform back from
the physical Hamiltonian $\tilde H_0$ in Eq.~\eqref{eq:h0tilde} to the original
Hamiltonian $H_0$, Eq.~\eqref{eq:h0}, with arbitrary $\lambda$ and $a$ as long
as Eq.~\eqref{eq:k0} is satisfied.

In order to investigate the dynamics of the full Hamiltonian $H$, 
the force term $-FxI$ may 
be added to the exact solution as a perturbation. In that
case one concludes that in the presence of spin-orbit coupling, the
two branches (effective spin components) respond to a force differently, 
as one of them accelerates faster than the other.

\subsection{Semiclassical solution}

We now treat $H$ including the effect of the force $F$ explicitly, and 
also carefully tracing the effects of Berry curvatures 
in this system. The full Zeeman term now is a position- and momentum-dependent 
matrix:
\be
\BdotSigma =
\frac{\hbar\rabi}{2}
\Big(
\cos\frac{x}{\lambda}\, \sigma_1 + \sin\frac{x}{\lambda}\, \sigma_{2}
\Big)
-\frac{\hbar^{2}}{m}ak\sigma_{3}
\,.
\ee
As this term contains both position and momentum operators,
it does not conform to either of the cases addressed in Sec.~\ref{sec:particular}.
We therefore have to resort to the semiclassical approach.

The direction of the vector $\vec B$ can be written in terms
of the spherical angles $\alpha(k)$ and $\phi(x)$ as introduced
in Eq.~\eqref{eq:spinor}. In particular, 
\bsa
\tan\alpha & =-\frac{\kappa}{k}\,,\\
\phi & =\frac{x}{\lambda}\,.
\esa 
The eigenvalues of the matrix $\BdotSigma$ are
\begin{equation}
\pm|\vec B|=\pm\frac{\hbar\rabi}{2\kappa}\sqrt{\kappa^{2}+k^{2}} \,.
\end{equation}
For adiabatic approximation to hold, it is sufficient that
the gap between the two branches, $2|\vec B|$, is larger than
all the other relevant energy scales. This condition in particular
enforces the single-minimum condition Eq.~\eqref{eq:oneMinimum}.
The Berry connections are $\mathcal{A}^{(k)}=0$ and
\be
\mathcal{A}^{(r)}=\mp\frac{k}{2\lambda\sqrt{\kappa^{2}+k^{2}}}
\ee
in this case.
Note that due to the nonvanishing ${\mathcal A}^{(r)}$, cf. Eq.~\eqref{eq:kc}, 
the physical momentum differs from the canonical momentum,
\be
k_{\mathrm{c}}=k\left(1\pm\frac{1}{2\lambda\sqrt{\kappa^{2}+k^{2}}}\right)
\,.
\ee

Position and momentum spaces remain flat, $\Theta^{(k,k)}=\Theta^{(r,r)}=0$,
and only phase-space Berry curvature is nonzero and opposite
for the two branches:
\be
\Theta^{(k,r)}=-\Theta^{(r,k)}
=\pm\frac{\kappa^{2}}{2\lambda(\kappa^{2}+k^{2})^{\frac{3}{2}}}
\,.
\ee
The scalar potentials \eqref{eq:PotDef} are $\mathcal{V}^{(k)}=0$ and
\begin{equation}
\mathcal{V}^{(r)}
=\frac{\hbar^{2}}{8m\lambda^{2}}
\frac{\kappa^{2}}{\kappa^{2}+k^{2}}\,,
\end{equation}
and thus the full potential $\mathcal{V}$ is 
\be
\mathcal{V} =
\pm\frac{\hbar\rabi}{2\kappa}\sqrt{\kappa^{2}+k^{2}}
+\frac{\hbar^{2}}{8m\lambda^{2}}\frac{\kappa^{2}}{\kappa^{2}+k^{2}}
\,.
\ee
Semiclassical dynamics follows the equations \eqref{eqs:SemiClassical}
\bsa
\dot k_\mathrm{c}  = & 
\frac{F}{\hbar} 
(1
-
\Theta^{(r,k)}
)
\,, \\
\dot {x}_{\mathrm{c}}  = &
\frac{\hbar}{m}{k}_{\mathrm{c}}
(1+\Theta^{(k,r)})
+\frac{1}{\hbar} \nabla^{(k)} \mathcal{V}
\,,
\esa
which in general are quite complicated. 

However, for our semiclassical approach to be valid, we should
keep only the lowest-order corrections in the small parameters
$1/\lambda a$ and $1/\lambda\kappa$. Furthermore, in order
to prevent the kinetic energy from causing sizeable interbranch
transitions, we limit ourselves to the $|k|\ll\kappa$ regime,
where adiabatic approximation holds. These approximations
lead to
\bsa
k_{\mathrm{c}}& \approx k\left(1\pm\frac{1}{2\lambda\kappa}\right)\,, \\
\Theta^{(r,k)} & = - \Theta^{(k,r)} \approx\mp\frac{1}{2\lambda\kappa}\,, \\
\frac{1}{\hbar}\nabla^{(k)} \mathcal{V} & 
\approx \pm\frac{k}{\kappa} 
\frac{\rabi}{2\kappa} 
\,. 
\esa
Therefore, equations of motion in this limit are
\bsa
\dot k_{\mathrm{c}} & =\frac{F}{\hbar}
\left(1\pm\frac{1}{2\lambda\kappa}\right)
\,,\\
\dot x_{\mathrm{c}} & =\frac{\hbar}{m}k_{\mathrm{c}}
\left(1\pm\frac{1}{2\lambda\kappa}\pm\frac{a}{\kappa}\right)
\,,
\esa
and for the center-of-mass motion we obtain the following closed equation:
\be
\label{eq:BerryClosedEOM}
\ddot x_\mathrm{c}=
\frac{F}{m} 
\left(1\pm\frac{1}{\kappa}\left(\frac{1}{\lambda}+a\right)\right)
\,.
\ee
From this equation one can read off the effective mass,
which is then seen to match the one given in Eq.~\eqref{eq:EffectiveMass}.

\begin{figure}[t]
\begin{center}
\includegraphics[width=1.68in]{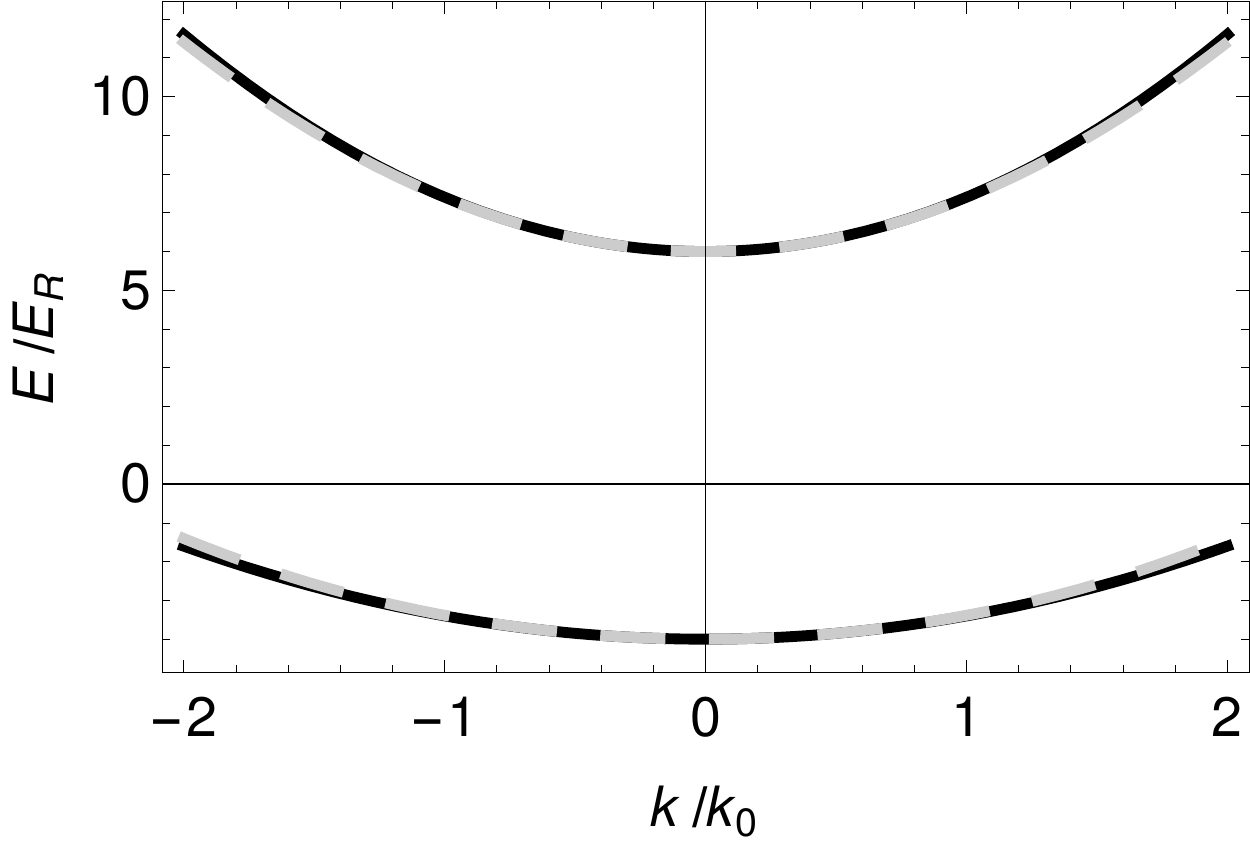}
\includegraphics[width=1.68in]{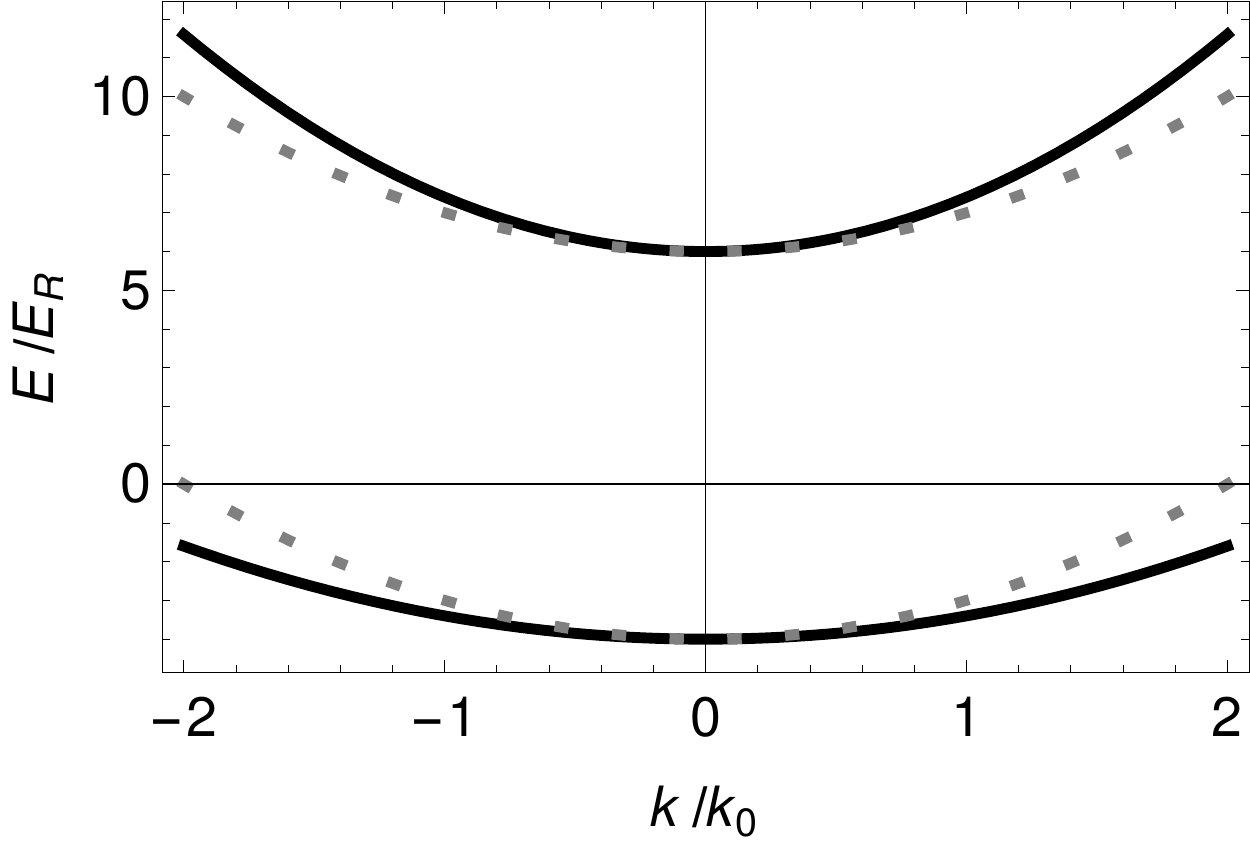}
\caption{Dispersion of the two branches in the units of the recoil
energy $E_R$ and the recoil wavevector $k_0$ in the semiclassical
single-minimum regime, $\hbar\rabi/E_R = 10$. 
The 
effective mass
from the phase-space Berry-curvature 
in Eq.~\eqref{eq:BerryClosedEOM} 
(solid line)
is compared to the exact result (dashed line, left) and the
bare-mass approximation $m^*_\pm = m$ (dotted line, right).}
\label{fig2}
\end{center}
\end{figure}

We conclude that the effective mass in this model 
in semiclassical approximation is correctly
captured for both branches by the phase-space Berry curvature.
The Berry-curvature result is compared with the bare mass and
also with the exact solution in Fig.~\ref{fig2}.
Thus, the effective mass measurement is a direct probe
of the phase-space Berry curvature in this system. Therefore,
effective-mass measurements in Refs.~\cite{LinEtAl2011b,ZhangEtAl2012}
can be reinterpreted as the
first measurements of phase-space Berry curvature
in ultracold-atom systems. This conclusion is the second
main result of the paper.

\section{Summary and outlook}
\label{sec:summary}

In summary, we have presented a derivation of equations of motion in the
presence of position-space, momentum-space and phase-space Berry curvatures. We
have only relied on the Hamiltonian formalism, and we have clearly separated
adiabatic approximation, semiclassical approximation, and the
low-curvature limit. 
Our approach has resulted in the Heisenberg equations of
motion Eqs.~\eqref{eqs:QuantumHeisenberg},
which are written down with no reference to semiclassical approximation.
From them, we have derived the semiclassical equations of motion
Eqs.~\eqref{eqs:SemiClassical}.  In the limit of
small curvature they reduce to
Eqs.~\eqref{eqs:SemiClassicalSmallCurv}.
The latter can be directly compared to the results of
Ref.~\cite{SundaramNiu1999}.
Moreover, we have investigated the
semiclassical dynamics in a system with an equal Rashba-Dresselhaus 
spin-orbit coupling system in a single spatial dimension. 
We have concluded that in the strong-coupling regime,
where the dispersions of both branches are quadratic, the effective mass
is directly related to phase-space Berry curvature.

Several directions of future work look promising. 
On the one hand, populating optical lattices with ultracold atoms one can go
far beyond the regimes readily achievable in solid-state physics. In particular,
it possible to engineer strong artificial magnetic fields \cite{AidelsburgerEtAl2013, 
MiyakeEtAl2013, AidelsburgerEtAl2015} as well as strong synthetic spin-orbit 
coupling \cite{LiviEtAl2016, WallEtAl2016, KolkowitzEtAl2017}, while motion of wave packets 
can be controlled and observed with remarkable precision 
\cite{AlbertiEtAl2009,AlbertiEtAl2010,HallerEtAl2010}. This motivates one to 
generalize the presented approach and revisit the effects of phase-space Berry 
curvature on a lattice. 
Furthermore, in higher-dimensional models, the effects of phase-space
Berry curvature most likely go beyond the rescaling of mass.
In the one-dimensional system we have investigated, the phase space is
two-dimensional, thus permitting only scalar curvature. In more dimensions,
this is no longer true.
Finally, in the semiclassical approach including the effects of 
harmonic confinement is straightforward, thus allowing one to consider
trapped systems and analyze their collective modes.

\begin{acknowledgments}
It is our pleasure to thank
Gediminas Juzeli\={u}nas,
Hannah Price,
Henk Stoof,
Marco Di Liberto,
Rembert Duine,
and
Viktor Novi\v{c}enko
for stimulating discussions.
This work was supported by the Research Council of
Lithuania under the Grant No.\ APP-4/2016.
J.~A. has received funding from the European Union's Horizon 2020 research and innovation
programme under the Marie Sk\l odowska-Curie grant agreement No 706839 (SPINSOCS).

\end{acknowledgments}

\bibliography{full}

\begin{thebibliography}{71}%
\makeatletter
\providecommand \@ifxundefined [1]{%
 \@ifx{#1\undefined}
}%
\providecommand \@ifnum [1]{%
 \ifnum #1\expandafter \@firstoftwo
 \else \expandafter \@secondoftwo
 \fi
}%
\providecommand \@ifx [1]{%
 \ifx #1\expandafter \@firstoftwo
 \else \expandafter \@secondoftwo
 \fi
}%
\providecommand \natexlab [1]{#1}%
\providecommand \enquote  [1]{``#1''}%
\providecommand \bibnamefont  [1]{#1}%
\providecommand \bibfnamefont [1]{#1}%
\providecommand \citenamefont [1]{#1}%
\providecommand \href@noop [0]{\@secondoftwo}%
\providecommand \href [0]{\begingroup \@sanitize@url \@href}%
\providecommand \@href[1]{\@@startlink{#1}\@@href}%
\providecommand \@@href[1]{\endgroup#1\@@endlink}%
\providecommand \@sanitize@url [0]{\catcode `\\12\catcode `\$12\catcode
  `\&12\catcode `\#12\catcode `\^12\catcode `\_12\catcode `\%12\relax}%
\providecommand \@@startlink[1]{}%
\providecommand \@@endlink[0]{}%
\providecommand \url  [0]{\begingroup\@sanitize@url \@url }%
\providecommand \@url [1]{\endgroup\@href {#1}{\urlprefix }}%
\providecommand \urlprefix  [0]{URL }%
\providecommand \Eprint [0]{\href }%
\providecommand \doibase [0]{http://dx.doi.org/}%
\providecommand \selectlanguage [0]{\@gobble}%
\providecommand \bibinfo  [0]{\@secondoftwo}%
\providecommand \bibfield  [0]{\@secondoftwo}%
\providecommand \translation [1]{[#1]}%
\providecommand \BibitemOpen [0]{}%
\providecommand \bibitemStop [0]{}%
\providecommand \bibitemNoStop [0]{.\EOS\space}%
\providecommand \EOS [0]{\spacefactor3000\relax}%
\providecommand \BibitemShut  [1]{\csname bibitem#1\endcsname}%
\let\auto@bib@innerbib\@empty
\bibitem [{\citenamefont {Frankel}(2004)}]{Frankel2004}%
  \BibitemOpen
  \bibfield  {author} {\bibinfo {author} {\bibfnamefont {T.}~\bibnamefont
  {Frankel}},\ }\href {\doibase 10.1017/CBO9781139061377} {\emph {\bibinfo
  {title} {The Geometry of Physics: An Introduction}}}\ (\bibinfo  {publisher}
  {Cambridge University Press},\ \bibinfo {year} {2004})\BibitemShut {NoStop}%
\bibitem [{\citenamefont {Einstein}(1915)}]{Einstein1915}%
  \BibitemOpen
  \bibfield  {author} {\bibinfo {author} {\bibfnamefont {A.}~\bibnamefont
  {Einstein}},\ }\bibfield  {title} {\enquote {\bibinfo {title} {{Die}
  {Feldgleichungen} der {Gravitation}},}\ }\href {\doibase
  10.1002/3527608958.ch5} {\bibfield  {journal} {\bibinfo  {journal}
  {Sitzungsber. Preuss. Akad. Wiss.}\ }\textbf {\bibinfo {volume} {25}},\
  \bibinfo {pages} {844--847} (\bibinfo {year} {1915})}\BibitemShut {NoStop}%
\bibitem [{\citenamefont {Yang}\ and\ \citenamefont
  {Mills}(1954)}]{YangMills1954}%
  \BibitemOpen
  \bibfield  {author} {\bibinfo {author} {\bibfnamefont {C.~N.}\ \bibnamefont
  {Yang}}\ and\ \bibinfo {author} {\bibfnamefont {R.~L.}\ \bibnamefont
  {Mills}},\ }\bibfield  {title} {\enquote {\bibinfo {title} {Conservation of
  isotopic spin and isotopic gauge invariance},}\ }\href {\doibase
  10.1103/PhysRev.96.191} {\bibfield  {journal} {\bibinfo  {journal} {Phys.
  Rev.}\ }\textbf {\bibinfo {volume} {96}},\ \bibinfo {pages} {191--195}
  (\bibinfo {year} {1954})}\BibitemShut {NoStop}%
\bibitem [{\citenamefont {Xiao}\ \emph {et~al.}(2010)\citenamefont {Xiao},
  \citenamefont {Chang},\ and\ \citenamefont {Niu}}]{XiaoEtAl2010}%
  \BibitemOpen
  \bibfield  {author} {\bibinfo {author} {\bibfnamefont {D.}~\bibnamefont
  {Xiao}}, \bibinfo {author} {\bibfnamefont {M.-C.}\ \bibnamefont {Chang}}, \
  and\ \bibinfo {author} {\bibfnamefont {Q.}~\bibnamefont {Niu}},\ }\bibfield
  {title} {\enquote {\bibinfo {title} {Berry phase effects on electronic
  properties},}\ }\href {\doibase 10.1103/RevModPhys.82.1959} {\bibfield
  {journal} {\bibinfo  {journal} {Rev. Mod. Phys.}\ }\textbf {\bibinfo {volume}
  {82}},\ \bibinfo {pages} {1959--2007} (\bibinfo {year} {2010})}\BibitemShut
  {NoStop}%
\bibitem [{\citenamefont {Fujita}\ \emph {et~al.}(2011)\citenamefont {Fujita},
  \citenamefont {Jalil}, \citenamefont {Tan},\ and\ \citenamefont
  {Murakami}}]{FujitaEtAl2011}%
  \BibitemOpen
  \bibfield  {author} {\bibinfo {author} {\bibfnamefont {T.}~\bibnamefont
  {Fujita}}, \bibinfo {author} {\bibfnamefont {M.~B.~A.}\ \bibnamefont
  {Jalil}}, \bibinfo {author} {\bibfnamefont {S.~G.}\ \bibnamefont {Tan}}, \
  and\ \bibinfo {author} {\bibfnamefont {S.}~\bibnamefont {Murakami}},\
  }\bibfield  {title} {\enquote {\bibinfo {title} {Gauge fields in
  spintronics},}\ }\href {\doibase 10.1063/1.3665219} {\bibfield  {journal}
  {\bibinfo  {journal} {J. Appl. Phys.}\ }\textbf {\bibinfo {volume} {110}},\
  \bibinfo {pages} {121301} (\bibinfo {year} {2011})}\BibitemShut {NoStop}%
\bibitem [{\citenamefont {Manchon}\ \emph {et~al.}(2015)\citenamefont
  {Manchon}, \citenamefont {Koo}, \citenamefont {Nitta}, \citenamefont
  {Frolov},\ and\ \citenamefont {Duine}}]{ManchonEtAl2015}%
  \BibitemOpen
  \bibfield  {author} {\bibinfo {author} {\bibfnamefont {A.}~\bibnamefont
  {Manchon}}, \bibinfo {author} {\bibfnamefont {H.~C.}\ \bibnamefont {Koo}},
  \bibinfo {author} {\bibfnamefont {J.}~\bibnamefont {Nitta}}, \bibinfo
  {author} {\bibfnamefont {S.~M.}\ \bibnamefont {Frolov}}, \ and\ \bibinfo
  {author} {\bibfnamefont {R.~A.}\ \bibnamefont {Duine}},\ }\bibfield  {title}
  {\enquote {\bibinfo {title} {New perspectives for {Rashba} spin-orbit
  coupling},}\ }\href {\doibase 10.1038/nmat4360} {\bibfield  {journal}
  {\bibinfo  {journal} {Nat. Mater.}\ }\textbf {\bibinfo {volume} {14}},\
  \bibinfo {pages} {871--882} (\bibinfo {year} {2015})}\BibitemShut {NoStop}%
\bibitem [{\citenamefont {Stuhl}\ \emph {et~al.}(2015)\citenamefont {Stuhl},
  \citenamefont {Lu}, \citenamefont {Aycock}, \citenamefont {Genkina},\ and\
  \citenamefont {Spielman}}]{StuhlEtAl2015}%
  \BibitemOpen
  \bibfield  {author} {\bibinfo {author} {\bibfnamefont {B.~K.}\ \bibnamefont
  {Stuhl}}, \bibinfo {author} {\bibfnamefont {H.-I.}\ \bibnamefont {Lu}},
  \bibinfo {author} {\bibfnamefont {L.~M.}\ \bibnamefont {Aycock}}, \bibinfo
  {author} {\bibfnamefont {D.}~\bibnamefont {Genkina}}, \ and\ \bibinfo
  {author} {\bibfnamefont {I.~B.}\ \bibnamefont {Spielman}},\ }\bibfield
  {title} {\enquote {\bibinfo {title} {Visualizing edge states with an atomic
  {B}ose gas in the quantum {H}all regime},}\ }\href {\doibase
  10.1126/science.aaa8515} {\bibfield  {journal} {\bibinfo  {journal}
  {Science}\ }\textbf {\bibinfo {volume} {349}},\ \bibinfo {pages} {1514--1518}
  (\bibinfo {year} {2015})}\BibitemShut {NoStop}%
\bibitem [{\citenamefont {Livi}\ \emph {et~al.}(2016)\citenamefont {Livi},
  \citenamefont {Cappellini}, \citenamefont {Diem}, \citenamefont {Franchi},
  \citenamefont {Clivati}, \citenamefont {Frittelli}, \citenamefont {Levi},
  \citenamefont {Calonico}, \citenamefont {Catani}, \citenamefont {Inguscio},\
  and\ \citenamefont {Fallani}}]{LiviEtAl2016}%
  \BibitemOpen
  \bibfield  {author} {\bibinfo {author} {\bibfnamefont {L.~F.}\ \bibnamefont
  {Livi}}, \bibinfo {author} {\bibfnamefont {G.}~\bibnamefont {Cappellini}},
  \bibinfo {author} {\bibfnamefont {M.}~\bibnamefont {Diem}}, \bibinfo {author}
  {\bibfnamefont {L.}~\bibnamefont {Franchi}}, \bibinfo {author} {\bibfnamefont
  {C.}~\bibnamefont {Clivati}}, \bibinfo {author} {\bibfnamefont
  {M.}~\bibnamefont {Frittelli}}, \bibinfo {author} {\bibfnamefont
  {F.}~\bibnamefont {Levi}}, \bibinfo {author} {\bibfnamefont {D.}~\bibnamefont
  {Calonico}}, \bibinfo {author} {\bibfnamefont {J.}~\bibnamefont {Catani}},
  \bibinfo {author} {\bibfnamefont {M.}~\bibnamefont {Inguscio}}, \ and\
  \bibinfo {author} {\bibfnamefont {L.}~\bibnamefont {Fallani}},\ }\bibfield
  {title} {\enquote {\bibinfo {title} {Synthetic dimensions and spin-orbit
  coupling with an optical clock transition},}\ }\href {\doibase
  10.1103/PhysRevLett.117.220401} {\bibfield  {journal} {\bibinfo  {journal}
  {Phys. Rev. Lett.}\ }\textbf {\bibinfo {volume} {117}},\ \bibinfo {pages}
  {220401} (\bibinfo {year} {2016})}\BibitemShut {NoStop}%
\bibitem [{\citenamefont {Wall}\ \emph {et~al.}(2016)\citenamefont {Wall},
  \citenamefont {Koller}, \citenamefont {Li}, \citenamefont {Zhang},
  \citenamefont {Cooper}, \citenamefont {Ye},\ and\ \citenamefont
  {Rey}}]{WallEtAl2016}%
  \BibitemOpen
  \bibfield  {author} {\bibinfo {author} {\bibfnamefont {M.~L.}\ \bibnamefont
  {Wall}}, \bibinfo {author} {\bibfnamefont {A.~P.}\ \bibnamefont {Koller}},
  \bibinfo {author} {\bibfnamefont {S.}~\bibnamefont {Li}}, \bibinfo {author}
  {\bibfnamefont {X.}~\bibnamefont {Zhang}}, \bibinfo {author} {\bibfnamefont
  {N.~R.}\ \bibnamefont {Cooper}}, \bibinfo {author} {\bibfnamefont
  {J.}~\bibnamefont {Ye}}, \ and\ \bibinfo {author} {\bibfnamefont {A.~M.}\
  \bibnamefont {Rey}},\ }\bibfield  {title} {\enquote {\bibinfo {title}
  {Synthetic spin-orbit coupling in an optical lattice clock},}\ }\href
  {\doibase 10.1103/PhysRevLett.116.035301} {\bibfield  {journal} {\bibinfo
  {journal} {Phys. Rev. Lett.}\ }\textbf {\bibinfo {volume} {116}},\ \bibinfo
  {pages} {035301} (\bibinfo {year} {2016})}\BibitemShut {NoStop}%
\bibitem [{\citenamefont {Dalibard}\ \emph {et~al.}(2011)\citenamefont
  {Dalibard}, \citenamefont {Gerbier}, \citenamefont {Juzeli\={u}nas},\ and\
  \citenamefont {\"{O}hberg}}]{DalibardEtAl2011}%
  \BibitemOpen
  \bibfield  {author} {\bibinfo {author} {\bibfnamefont {J.}~\bibnamefont
  {Dalibard}}, \bibinfo {author} {\bibfnamefont {F.}~\bibnamefont {Gerbier}},
  \bibinfo {author} {\bibfnamefont {G.}~\bibnamefont {Juzeli\={u}nas}}, \ and\
  \bibinfo {author} {\bibfnamefont {P.}~\bibnamefont {\"{O}hberg}},\ }\bibfield
   {title} {\enquote {\bibinfo {title} {Colloquium: {A}rtificial gauge
  potentials for neutral atoms},}\ }\href {\doibase 10.1103/RevModPhys.83.1523}
  {\bibfield  {journal} {\bibinfo  {journal} {Rev. Mod. Phys.}\ }\textbf
  {\bibinfo {volume} {83}},\ \bibinfo {pages} {1523--1543} (\bibinfo {year}
  {2011})}\BibitemShut {NoStop}%
\bibitem [{\citenamefont {Galitski}\ and\ \citenamefont
  {Spielman}(2013)}]{GalitskiSpielm2013}%
  \BibitemOpen
  \bibfield  {author} {\bibinfo {author} {\bibfnamefont {V.}~\bibnamefont
  {Galitski}}\ and\ \bibinfo {author} {\bibfnamefont {I.~B.}\ \bibnamefont
  {Spielman}},\ }\bibfield  {title} {\enquote {\bibinfo {title} {Spin-orbit
  coupling in quantum gases},}\ }\href {\doibase 10.1038/nature11841}
  {\bibfield  {journal} {\bibinfo  {journal} {Nature}\ }\textbf {\bibinfo
  {volume} {494}},\ \bibinfo {pages} {49--54} (\bibinfo {year}
  {2013})}\BibitemShut {NoStop}%
\bibitem [{\citenamefont {Goldman}\ \emph {et~al.}(2014)\citenamefont
  {Goldman}, \citenamefont {Juzeli{\=u}nas}, \citenamefont {\"{O}hberg},\ and\
  \citenamefont {Spielman}}]{GoldmanEtAl2014}%
  \BibitemOpen
  \bibfield  {author} {\bibinfo {author} {\bibfnamefont {N.}~\bibnamefont
  {Goldman}}, \bibinfo {author} {\bibfnamefont {G.}~\bibnamefont
  {Juzeli{\=u}nas}}, \bibinfo {author} {\bibfnamefont {P.}~\bibnamefont
  {\"{O}hberg}}, \ and\ \bibinfo {author} {\bibfnamefont {I.~B.}\ \bibnamefont
  {Spielman}},\ }\bibfield  {title} {\enquote {\bibinfo {title} {Light-induced
  gauge fields for ultracold atoms},}\ }\href {\doibase
  10.1088/0034-4885/77/12/126401} {\bibfield  {journal} {\bibinfo  {journal}
  {Rep. Prog. Phys.}\ }\textbf {\bibinfo {volume} {77}},\ \bibinfo {pages}
  {126401} (\bibinfo {year} {2014})}\BibitemShut {NoStop}%
\bibitem [{\citenamefont {Zhai}(2015)}]{Zhai2015}%
  \BibitemOpen
  \bibfield  {author} {\bibinfo {author} {\bibfnamefont {H.}~\bibnamefont
  {Zhai}},\ }\bibfield  {title} {\enquote {\bibinfo {title} {Degenerate quantum
  gases with spin--orbit coupling: a review},}\ }\href {\doibase
  10.1088/0034-4885/78/2/026001} {\bibfield  {journal} {\bibinfo  {journal}
  {Rep. Prog. Phys.}\ }\textbf {\bibinfo {volume} {78}},\ \bibinfo {pages}
  {026001} (\bibinfo {year} {2015})}\BibitemShut {NoStop}%
\bibitem [{\citenamefont {Rashba}(1960)}]{Rashba1960}%
  \BibitemOpen
  \bibfield  {author} {\bibinfo {author} {\bibfnamefont {E.}~\bibnamefont
  {Rashba}},\ }\bibfield  {title} {\enquote {\bibinfo {title} {Properties of
  semiconductors with an extremum loop. 1. {Cyclotron} and combinational
  resonance in a magnetic field perpendicular to the plane of the loop},}\
  }\href@noop {} {\bibfield  {journal} {\bibinfo  {journal} {Sov. Phys. Solid
  State}\ }\textbf {\bibinfo {volume} {2}},\ \bibinfo {pages} {1109--1122}
  (\bibinfo {year} {1960})}\BibitemShut {NoStop}%
\bibitem [{\citenamefont {Dresselhaus}(1955)}]{Dresselhaus1955}%
  \BibitemOpen
  \bibfield  {author} {\bibinfo {author} {\bibfnamefont {G.}~\bibnamefont
  {Dresselhaus}},\ }\bibfield  {title} {\enquote {\bibinfo {title}
  {{Spin}-{Orbit} {Coupling} {Effects} in {Zinc} {Blende} {Structures}},}\
  }\href {\doibase 10.1103/PhysRev.100.580} {\bibfield  {journal} {\bibinfo
  {journal} {Phys. Rev.}\ }\textbf {\bibinfo {volume} {100}},\ \bibinfo {pages}
  {580--586} (\bibinfo {year} {1955})}\BibitemShut {NoStop}%
\bibitem [{\citenamefont {Lin}\ \emph {et~al.}(2011{\natexlab{a}})\citenamefont
  {Lin}, \citenamefont {{Jim{\'e}nez-Garc{\'{\i}}a}},\ and\ \citenamefont
  {Spielman}}]{LinEtAl2011a}%
  \BibitemOpen
  \bibfield  {author} {\bibinfo {author} {\bibfnamefont {Y.-J.}\ \bibnamefont
  {Lin}}, \bibinfo {author} {\bibfnamefont {K.}~\bibnamefont
  {{Jim{\'e}nez-Garc{\'{\i}}a}}}, \ and\ \bibinfo {author} {\bibfnamefont
  {I.~B.}\ \bibnamefont {Spielman}},\ }\bibfield  {title} {\enquote {\bibinfo
  {title} {Spin-orbit-coupled {B}ose-{E}instein condensates},}\ }\href
  {\doibase 10.1038/nature09887} {\bibfield  {journal} {\bibinfo  {journal}
  {Nature}\ }\textbf {\bibinfo {volume} {471}},\ \bibinfo {pages} {83--86}
  (\bibinfo {year} {2011}{\natexlab{a}})}\BibitemShut {NoStop}%
\bibitem [{\citenamefont {Li}\ \emph {et~al.}(2015)\citenamefont {Li},
  \citenamefont {Martone},\ and\ \citenamefont {Stringari}}]{LiEtAl2015}%
  \BibitemOpen
  \bibfield  {author} {\bibinfo {author} {\bibfnamefont {Y.}~\bibnamefont
  {Li}}, \bibinfo {author} {\bibfnamefont {G.~I.}\ \bibnamefont {Martone}}, \
  and\ \bibinfo {author} {\bibfnamefont {S.}~\bibnamefont {Stringari}},\
  }\enquote {\bibinfo {title} {Spin-orbit coupled {Bose}-{Einstein}
  condensates},}\ in\ \href {\doibase 10.1142/9789814667746_0005} {\emph
  {\bibinfo {booktitle} {Annual Review of Cold Atoms and Molecules}}}\
  (\bibinfo  {publisher} {World Scientific},\ \bibinfo {year} {2015})\
  Chap.~\bibinfo {chapter} {5}, pp.\ \bibinfo {pages} {201--250}\BibitemShut
  {NoStop}%
\bibitem [{\citenamefont {Campbell}\ and\ \citenamefont
  {Spielman}(2016)}]{CampbellSpielman2016}%
  \BibitemOpen
  \bibfield  {author} {\bibinfo {author} {\bibfnamefont {D.~L.}\ \bibnamefont
  {Campbell}}\ and\ \bibinfo {author} {\bibfnamefont {I.~B.}\ \bibnamefont
  {Spielman}},\ }\bibfield  {title} {\enquote {\bibinfo {title} {Rashba
  realization: {R}aman with {RF}},}\ }\href
  {http://stacks.iop.org/1367-2630/18/i=3/a=033035} {\bibfield  {journal}
  {\bibinfo  {journal} {New J. Phys.}\ }\textbf {\bibinfo {volume} {18}},\
  \bibinfo {pages} {033035} (\bibinfo {year} {2016})}\BibitemShut {NoStop}%
\bibitem [{\citenamefont {Huang}\ \emph {et~al.}(2016)\citenamefont {Huang},
  \citenamefont {Meng}, \citenamefont {Wang}, \citenamefont {Peng},
  \citenamefont {Zhang}, \citenamefont {Chen}, \citenamefont {Li},
  \citenamefont {Zhou},\ and\ \citenamefont {Zhang}}]{HuangEtAl2016}%
  \BibitemOpen
  \bibfield  {author} {\bibinfo {author} {\bibfnamefont {L.}~\bibnamefont
  {Huang}}, \bibinfo {author} {\bibfnamefont {Z.}~\bibnamefont {Meng}},
  \bibinfo {author} {\bibfnamefont {P.}~\bibnamefont {Wang}}, \bibinfo {author}
  {\bibfnamefont {P.}~\bibnamefont {Peng}}, \bibinfo {author} {\bibfnamefont
  {S.-L.}\ \bibnamefont {Zhang}}, \bibinfo {author} {\bibfnamefont
  {L.}~\bibnamefont {Chen}}, \bibinfo {author} {\bibfnamefont {D.}~\bibnamefont
  {Li}}, \bibinfo {author} {\bibfnamefont {Q.}~\bibnamefont {Zhou}}, \ and\
  \bibinfo {author} {\bibfnamefont {J.}~\bibnamefont {Zhang}},\ }\bibfield
  {title} {\enquote {\bibinfo {title} {{Experimental} realization of
  two-dimensional synthetic spin-orbit coupling in ultracold {Fermi} gases},}\
  }\href {\doibase 10.1038/nphys3672} {\bibfield  {journal} {\bibinfo
  {journal} {Nature Phys.}\ }\textbf {\bibinfo {volume} {12}},\ \bibinfo
  {pages} {540--544} (\bibinfo {year} {2016})}\BibitemShut {NoStop}%
\bibitem [{\citenamefont {Anderson}\ \emph {et~al.}(2012)\citenamefont
  {Anderson}, \citenamefont {Juzeli\ifmmode~\bar{u}\else \={u}\fi{}nas},
  \citenamefont {Galitski},\ and\ \citenamefont {Spielman}}]{AndersonEtAl2012}%
  \BibitemOpen
  \bibfield  {author} {\bibinfo {author} {\bibfnamefont {B.~M.}\ \bibnamefont
  {Anderson}}, \bibinfo {author} {\bibfnamefont {G.}~\bibnamefont
  {Juzeli\ifmmode~\bar{u}\else \={u}\fi{}nas}}, \bibinfo {author}
  {\bibfnamefont {V.~M.}\ \bibnamefont {Galitski}}, \ and\ \bibinfo {author}
  {\bibfnamefont {I.~B.}\ \bibnamefont {Spielman}},\ }\bibfield  {title}
  {\enquote {\bibinfo {title} {Synthetic 3d spin-orbit coupling},}\ }\href
  {\doibase 10.1103/PhysRevLett.108.235301} {\bibfield  {journal} {\bibinfo
  {journal} {Phys. Rev. Lett.}\ }\textbf {\bibinfo {volume} {108}},\ \bibinfo
  {pages} {235301} (\bibinfo {year} {2012})}\BibitemShut {NoStop}%
\bibitem [{\citenamefont {Anderson}\ \emph {et~al.}(2013)\citenamefont
  {Anderson}, \citenamefont {Spielman},\ and\ \citenamefont
  {Juzeli{\=u}nas}}]{AndersonEtAl2013}%
  \BibitemOpen
  \bibfield  {author} {\bibinfo {author} {\bibfnamefont {B.~M.}\ \bibnamefont
  {Anderson}}, \bibinfo {author} {\bibfnamefont {I.~B.}\ \bibnamefont
  {Spielman}}, \ and\ \bibinfo {author} {\bibfnamefont {G.}~\bibnamefont
  {Juzeli{\=u}nas}},\ }\bibfield  {title} {\enquote {\bibinfo {title}
  {Magnetically generated spin-orbit coupling for ultracold atoms},}\ }\href
  {\doibase 10.1103/PhysRevLett.111.125301} {\bibfield  {journal} {\bibinfo
  {journal} {Phys. Rev. Lett.}\ }\textbf {\bibinfo {volume} {111}},\ \bibinfo
  {pages} {125301} (\bibinfo {year} {2013})}\BibitemShut {NoStop}%
\bibitem [{\citenamefont {Dub\ifmmode~\check{c}\else \v{c}\fi{}ek}\ \emph
  {et~al.}(2015)\citenamefont {Dub\ifmmode~\check{c}\else \v{c}\fi{}ek},
  \citenamefont {Kennedy}, \citenamefont {Lu}, \citenamefont {Ketterle},
  \citenamefont {Solja\ifmmode \check{c}\else
  \v{c}\fi{}i\ifmmode~\acute{c}\else \'{c}\fi{}},\ and\ \citenamefont
  {Buljan}}]{DubcekEtAl2015}%
  \BibitemOpen
  \bibfield  {author} {\bibinfo {author} {\bibfnamefont {T.}~\bibnamefont
  {Dub\ifmmode~\check{c}\else \v{c}\fi{}ek}}, \bibinfo {author} {\bibfnamefont
  {C.~J.}\ \bibnamefont {Kennedy}}, \bibinfo {author} {\bibfnamefont
  {L.}~\bibnamefont {Lu}}, \bibinfo {author} {\bibfnamefont {W.}~\bibnamefont
  {Ketterle}}, \bibinfo {author} {\bibfnamefont {M.}~\bibnamefont
  {Solja\ifmmode \check{c}\else \v{c}\fi{}i\ifmmode~\acute{c}\else
  \'{c}\fi{}}}, \ and\ \bibinfo {author} {\bibfnamefont {H.}~\bibnamefont
  {Buljan}},\ }\bibfield  {title} {\enquote {\bibinfo {title} {Weyl points in
  three-dimensional optical lattices: Synthetic magnetic monopoles in momentum
  space},}\ }\href {\doibase 10.1103/PhysRevLett.114.225301} {\bibfield
  {journal} {\bibinfo  {journal} {Phys. Rev. Lett.}\ }\textbf {\bibinfo
  {volume} {114}},\ \bibinfo {pages} {225301} (\bibinfo {year}
  {2015})}\BibitemShut {NoStop}%
\bibitem [{\citenamefont {Armaitis}\ \emph {et~al.}(2017)\citenamefont
  {Armaitis}, \citenamefont {Ruseckas},\ and\ \citenamefont
  {Juzeli\ifmmode~\bar{u}\else \={u}\fi{}nas}}]{ArmaitisEtAl2017}%
  \BibitemOpen
  \bibfield  {author} {\bibinfo {author} {\bibfnamefont {J.}~\bibnamefont
  {Armaitis}}, \bibinfo {author} {\bibfnamefont {J.}~\bibnamefont {Ruseckas}},
  \ and\ \bibinfo {author} {\bibfnamefont {G.}~\bibnamefont
  {Juzeli\ifmmode~\bar{u}\else \={u}\fi{}nas}},\ }\bibfield  {title} {\enquote
  {\bibinfo {title} {Omnidirectional spin {Hall} effect in a {Weyl}
  spin-orbit-coupled atomic gas},}\ }\href {\doibase
  10.1103/PhysRevA.95.033635} {\bibfield  {journal} {\bibinfo  {journal} {Phys.
  Rev. A}\ }\textbf {\bibinfo {volume} {95}},\ \bibinfo {pages} {033635}
  (\bibinfo {year} {2017})}\BibitemShut {NoStop}%
\bibitem [{\citenamefont {Struck}\ \emph {et~al.}(2012)\citenamefont {Struck},
  \citenamefont {\"{O}lschl\"{a}ger}, \citenamefont {Weinberg}, \citenamefont
  {Hauke}, \citenamefont {Simonet}, \citenamefont {Eckardt}, \citenamefont
  {Lewenstein}, \citenamefont {Sengstock},\ and\ \citenamefont
  {Windpassinger}}]{StruckEtAl2012}%
  \BibitemOpen
  \bibfield  {author} {\bibinfo {author} {\bibfnamefont {J.}~\bibnamefont
  {Struck}}, \bibinfo {author} {\bibfnamefont {C.}~\bibnamefont
  {\"{O}lschl\"{a}ger}}, \bibinfo {author} {\bibfnamefont {M.}~\bibnamefont
  {Weinberg}}, \bibinfo {author} {\bibfnamefont {P.}~\bibnamefont {Hauke}},
  \bibinfo {author} {\bibfnamefont {J.}~\bibnamefont {Simonet}}, \bibinfo
  {author} {\bibfnamefont {A.}~\bibnamefont {Eckardt}}, \bibinfo {author}
  {\bibfnamefont {M.}~\bibnamefont {Lewenstein}}, \bibinfo {author}
  {\bibfnamefont {K.}~\bibnamefont {Sengstock}}, \ and\ \bibinfo {author}
  {\bibfnamefont {P.}~\bibnamefont {Windpassinger}},\ }\bibfield  {title}
  {\enquote {\bibinfo {title} {Tunable gauge potential for neutral and spinless
  particles in driven optical lattices},}\ }\href {\doibase
  10.1103/PhysRevLett.108.225304} {\bibfield  {journal} {\bibinfo  {journal}
  {Phys. Rev. Lett.}\ }\textbf {\bibinfo {volume} {108}},\ \bibinfo {pages}
  {225304} (\bibinfo {year} {2012})}\BibitemShut {NoStop}%
\bibitem [{\citenamefont {Struck}\ \emph {et~al.}(2014)\citenamefont {Struck},
  \citenamefont {Simonet},\ and\ \citenamefont {Sengstock}}]{StruckEtAl2014}%
  \BibitemOpen
  \bibfield  {author} {\bibinfo {author} {\bibfnamefont {J.}~\bibnamefont
  {Struck}}, \bibinfo {author} {\bibfnamefont {J.}~\bibnamefont {Simonet}}, \
  and\ \bibinfo {author} {\bibfnamefont {K.}~\bibnamefont {Sengstock}},\
  }\bibfield  {title} {\enquote {\bibinfo {title} {Spin-orbit coupling in
  periodically driven optical lattices},}\ }\href {\doibase
  10.1103/PhysRevA.90.031601} {\bibfield  {journal} {\bibinfo  {journal} {Phys.
  Rev. A}\ }\textbf {\bibinfo {volume} {90}},\ \bibinfo {pages} {031601(R)}
  (\bibinfo {year} {2014})}\BibitemShut {NoStop}%
\bibitem [{\citenamefont {Eckardt}(2017)}]{Eckardt2016}%
  \BibitemOpen
  \bibfield  {author} {\bibinfo {author} {\bibfnamefont {Andr\'e}\ \bibnamefont
  {Eckardt}},\ }\bibfield  {title} {\enquote {\bibinfo {title} {Colloquium:
  Atomic quantum gases in periodically driven optical lattices},}\ }\href
  {\doibase 10.1103/RevModPhys.89.011004} {\bibfield  {journal} {\bibinfo
  {journal} {Rev. Mod. Phys.}\ }\textbf {\bibinfo {volume} {89}},\ \bibinfo
  {pages} {011004} (\bibinfo {year} {2017})}\BibitemShut {NoStop}%
\bibitem [{\citenamefont {Struck}\ \emph {et~al.}(2011)\citenamefont {Struck},
  \citenamefont {\"Olschl\"ager}, \citenamefont {{Le Targat}}, \citenamefont
  {{Soltan-Panahi}}, \citenamefont {Eckardt}, \citenamefont {Lewenstein},
  \citenamefont {Windpassinger},\ and\ \citenamefont
  {Sengstock}}]{StruckEtAl2011}%
  \BibitemOpen
  \bibfield  {author} {\bibinfo {author} {\bibfnamefont {J.}~\bibnamefont
  {Struck}}, \bibinfo {author} {\bibfnamefont {C.}~\bibnamefont
  {\"Olschl\"ager}}, \bibinfo {author} {\bibfnamefont {R.}~\bibnamefont {{Le
  Targat}}}, \bibinfo {author} {\bibfnamefont {P.}~\bibnamefont
  {{Soltan-Panahi}}}, \bibinfo {author} {\bibfnamefont {A.}~\bibnamefont
  {Eckardt}}, \bibinfo {author} {\bibfnamefont {M.}~\bibnamefont {Lewenstein}},
  \bibinfo {author} {\bibfnamefont {P.}~\bibnamefont {Windpassinger}}, \ and\
  \bibinfo {author} {\bibfnamefont {K.}~\bibnamefont {Sengstock}},\ }\bibfield
  {title} {\enquote {\bibinfo {title} {Quantum simulation of frustrated
  classical magnetism in triangular optical lattices},}\ }\href {\doibase
  10.1126/science.1207239} {\bibfield  {journal} {\bibinfo  {journal}
  {Science}\ }\textbf {\bibinfo {volume} {333}},\ \bibinfo {pages} {996--999}
  (\bibinfo {year} {2011})}\BibitemShut {NoStop}%
\bibitem [{\citenamefont {Lewenstein}\ \emph {et~al.}(2012)\citenamefont
  {Lewenstein}, \citenamefont {Sanpera},\ and\ \citenamefont
  {Ahufinger}}]{LewensteinEtAl2012}%
  \BibitemOpen
  \bibfield  {author} {\bibinfo {author} {\bibfnamefont {M.}~\bibnamefont
  {Lewenstein}}, \bibinfo {author} {\bibfnamefont {A.}~\bibnamefont {Sanpera}},
  \ and\ \bibinfo {author} {\bibfnamefont {V.}~\bibnamefont {Ahufinger}},\
  }\href
  {https://www.amazon.com/Ultracold-Atoms-Optical-Lattices-Simulating/dp/0199573123}
  {\emph {\bibinfo {title} {Ultracold atoms in optical lattices: {S}imulating
  quantum many-body systems}}}\ (\bibinfo  {publisher} {Oxford University
  Press},\ \bibinfo {year} {2012})\BibitemShut {NoStop}%
\bibitem [{\citenamefont {Struck}\ \emph {et~al.}(2013)\citenamefont {Struck},
  \citenamefont {Weinberg}, \citenamefont {\"Olschl\"ager}, \citenamefont
  {Windpassinger}, \citenamefont {Simonet}, \citenamefont {Sengstock},
  \citenamefont {H\"oppner}, \citenamefont {Hauke}, \citenamefont {Eckardt},
  \citenamefont {Lewenstein},\ and\ \citenamefont {Mathey}}]{StruckEtAl2013}%
  \BibitemOpen
  \bibfield  {author} {\bibinfo {author} {\bibfnamefont {J.}~\bibnamefont
  {Struck}}, \bibinfo {author} {\bibfnamefont {M.}~\bibnamefont {Weinberg}},
  \bibinfo {author} {\bibfnamefont {C.}~\bibnamefont {\"Olschl\"ager}},
  \bibinfo {author} {\bibfnamefont {P.}~\bibnamefont {Windpassinger}}, \bibinfo
  {author} {\bibfnamefont {J.}~\bibnamefont {Simonet}}, \bibinfo {author}
  {\bibfnamefont {K.}~\bibnamefont {Sengstock}}, \bibinfo {author}
  {\bibfnamefont {R.}~\bibnamefont {H\"oppner}}, \bibinfo {author}
  {\bibfnamefont {P.}~\bibnamefont {Hauke}}, \bibinfo {author} {\bibfnamefont
  {A.}~\bibnamefont {Eckardt}}, \bibinfo {author} {\bibfnamefont
  {M.}~\bibnamefont {Lewenstein}}, \ and\ \bibinfo {author} {\bibfnamefont
  {L.}~\bibnamefont {Mathey}},\ }\bibfield  {title} {\enquote {\bibinfo {title}
  {Engineering {Ising-XY} spin models in a triangular lattice via tunable
  artificial gauge fields},}\ }\href {\doibase 10.1038/nphys2750} {\bibfield
  {journal} {\bibinfo  {journal} {Nat. Phys.}\ }\textbf {\bibinfo {volume}
  {9}},\ \bibinfo {pages} {738--743} (\bibinfo {year} {2013})}\BibitemShut
  {NoStop}%
\bibitem [{\citenamefont {Aidelsburger}\ \emph {et~al.}(2013)\citenamefont
  {Aidelsburger}, \citenamefont {Atala}, \citenamefont {Lohse}, \citenamefont
  {Barreiro}, \citenamefont {Paredes},\ and\ \citenamefont
  {Bloch}}]{AidelsburgerEtAl2013}%
  \BibitemOpen
  \bibfield  {author} {\bibinfo {author} {\bibfnamefont {M.}~\bibnamefont
  {Aidelsburger}}, \bibinfo {author} {\bibfnamefont {M.}~\bibnamefont {Atala}},
  \bibinfo {author} {\bibfnamefont {M.}~\bibnamefont {Lohse}}, \bibinfo
  {author} {\bibfnamefont {J.~T.}\ \bibnamefont {Barreiro}}, \bibinfo {author}
  {\bibfnamefont {B.}~\bibnamefont {Paredes}}, \ and\ \bibinfo {author}
  {\bibfnamefont {I.}~\bibnamefont {Bloch}},\ }\bibfield  {title} {\enquote
  {\bibinfo {title} {Realization of the {H}ofstadter {H}amiltonian with
  ultracold atoms in optical lattices},}\ }\href {\doibase
  10.1103/PhysRevLett.111.185301} {\bibfield  {journal} {\bibinfo  {journal}
  {Phys. Rev. Lett.}\ }\textbf {\bibinfo {volume} {111}},\ \bibinfo {pages}
  {185301} (\bibinfo {year} {2013})}\BibitemShut {NoStop}%
\bibitem [{\citenamefont {Miyake}\ \emph {et~al.}(2013)\citenamefont {Miyake},
  \citenamefont {Siviloglou}, \citenamefont {Kennedy}, \citenamefont {Burton},\
  and\ \citenamefont {Ketterle}}]{MiyakeEtAl2013}%
  \BibitemOpen
  \bibfield  {author} {\bibinfo {author} {\bibfnamefont {H.}~\bibnamefont
  {Miyake}}, \bibinfo {author} {\bibfnamefont {G.~A.}\ \bibnamefont
  {Siviloglou}}, \bibinfo {author} {\bibfnamefont {C.~J.}\ \bibnamefont
  {Kennedy}}, \bibinfo {author} {\bibfnamefont {W.~C.}\ \bibnamefont {Burton}},
  \ and\ \bibinfo {author} {\bibfnamefont {W.}~\bibnamefont {Ketterle}},\
  }\bibfield  {title} {\enquote {\bibinfo {title} {Realizing the {H}arper
  {H}amiltonian with laser-assisted tunneling in optical lattices},}\ }\href
  {\doibase 10.1103/PhysRevLett.111.185302} {\bibfield  {journal} {\bibinfo
  {journal} {Phys. Rev. Lett.}\ }\textbf {\bibinfo {volume} {111}},\ \bibinfo
  {pages} {185302} (\bibinfo {year} {2013})}\BibitemShut {NoStop}%
\bibitem [{\citenamefont {Goldman}\ \emph {et~al.}(2016)\citenamefont
  {Goldman}, \citenamefont {Budich},\ and\ \citenamefont
  {Zoller}}]{GoldmanEtAl2016}%
  \BibitemOpen
  \bibfield  {author} {\bibinfo {author} {\bibfnamefont {N.}~\bibnamefont
  {Goldman}}, \bibinfo {author} {\bibfnamefont {J.~C.}\ \bibnamefont {Budich}},
  \ and\ \bibinfo {author} {\bibfnamefont {P.}~\bibnamefont {Zoller}},\
  }\bibfield  {title} {\enquote {\bibinfo {title} {Topological quantum matter
  with ultracold gases in optical lattices},}\ }\href {\doibase
  10.1038/nphys3803} {\bibfield  {journal} {\bibinfo  {journal} {Nat. Phys.}\
  }\textbf {\bibinfo {volume} {12}},\ \bibinfo {pages} {639--645} (\bibinfo
  {year} {2016})}\BibitemShut {NoStop}%
\bibitem [{\citenamefont {Celi}\ \emph {et~al.}(2014)\citenamefont {Celi},
  \citenamefont {Massignan}, \citenamefont {Ruseckas}, \citenamefont {Goldman},
  \citenamefont {Spielman}, \citenamefont {Juzeli\={u}nas},\ and\ \citenamefont
  {Lewenstein}}]{CeliEtAl2014}%
  \BibitemOpen
  \bibfield  {author} {\bibinfo {author} {\bibfnamefont {A.}~\bibnamefont
  {Celi}}, \bibinfo {author} {\bibfnamefont {P.}~\bibnamefont {Massignan}},
  \bibinfo {author} {\bibfnamefont {J.}~\bibnamefont {Ruseckas}}, \bibinfo
  {author} {\bibfnamefont {N.}~\bibnamefont {Goldman}}, \bibinfo {author}
  {\bibfnamefont {I.~B.}\ \bibnamefont {Spielman}}, \bibinfo {author}
  {\bibfnamefont {G.}~\bibnamefont {Juzeli\={u}nas}}, \ and\ \bibinfo {author}
  {\bibfnamefont {M.}~\bibnamefont {Lewenstein}},\ }\bibfield  {title}
  {\enquote {\bibinfo {title} {Synthetic gauge fields in synthetic
  dimensions},}\ }\href {\doibase 10.1103/PhysRevLett.112.043001} {\bibfield
  {journal} {\bibinfo  {journal} {Phys. Rev. Lett.}\ }\textbf {\bibinfo
  {volume} {112}},\ \bibinfo {pages} {043001} (\bibinfo {year}
  {2014})}\BibitemShut {NoStop}%
\bibitem [{\citenamefont {Mancini}\ \emph {et~al.}(2015)\citenamefont
  {Mancini}, \citenamefont {Pagano}, \citenamefont {Cappellini}, \citenamefont
  {Livi}, \citenamefont {Rider}, \citenamefont {Catani}, \citenamefont {Sias},
  \citenamefont {Zoller}, \citenamefont {Inguscio}, \citenamefont {Dalmonte},\
  and\ \citenamefont {Fallani}}]{ManciniEtAl2015}%
  \BibitemOpen
  \bibfield  {author} {\bibinfo {author} {\bibfnamefont {M.}~\bibnamefont
  {Mancini}}, \bibinfo {author} {\bibfnamefont {G.}~\bibnamefont {Pagano}},
  \bibinfo {author} {\bibfnamefont {G.}~\bibnamefont {Cappellini}}, \bibinfo
  {author} {\bibfnamefont {L.}~\bibnamefont {Livi}}, \bibinfo {author}
  {\bibfnamefont {M.}~\bibnamefont {Rider}}, \bibinfo {author} {\bibfnamefont
  {J.}~\bibnamefont {Catani}}, \bibinfo {author} {\bibfnamefont
  {C.}~\bibnamefont {Sias}}, \bibinfo {author} {\bibfnamefont {P.}~\bibnamefont
  {Zoller}}, \bibinfo {author} {\bibfnamefont {M.}~\bibnamefont {Inguscio}},
  \bibinfo {author} {\bibfnamefont {M.}~\bibnamefont {Dalmonte}}, \ and\
  \bibinfo {author} {\bibfnamefont {L.}~\bibnamefont {Fallani}},\ }\bibfield
  {title} {\enquote {\bibinfo {title} {Observation of chiral edge states with
  neutral fermions in synthetic {H}all ribbons},}\ }\href {\doibase
  10.1126/science.aaa8736} {\bibfield  {journal} {\bibinfo  {journal}
  {Science}\ }\textbf {\bibinfo {volume} {349}},\ \bibinfo {pages} {1510--1513}
  (\bibinfo {year} {2015})}\BibitemShut {NoStop}%
\bibitem [{\citenamefont {Price}\ \emph
  {et~al.}(2015{\natexlab{a}})\citenamefont {Price}, \citenamefont
  {Zilberberg}, \citenamefont {Ozawa}, \citenamefont {Carusotto},\ and\
  \citenamefont {Goldman}}]{PriceEtAl2015a}%
  \BibitemOpen
  \bibfield  {author} {\bibinfo {author} {\bibfnamefont {H.~M.}\ \bibnamefont
  {Price}}, \bibinfo {author} {\bibfnamefont {O.}~\bibnamefont {Zilberberg}},
  \bibinfo {author} {\bibfnamefont {T.}~\bibnamefont {Ozawa}}, \bibinfo
  {author} {\bibfnamefont {I.}~\bibnamefont {Carusotto}}, \ and\ \bibinfo
  {author} {\bibfnamefont {N.}~\bibnamefont {Goldman}},\ }\bibfield  {title}
  {\enquote {\bibinfo {title} {Four-dimensional quantum {H}all effect with
  ultracold atoms},}\ }\href {\doibase 10.1103/PhysRevLett.115.195303}
  {\bibfield  {journal} {\bibinfo  {journal} {Phys. Rev. Lett.}\ }\textbf
  {\bibinfo {volume} {115}},\ \bibinfo {pages} {195303} (\bibinfo {year}
  {2015}{\natexlab{a}})}\BibitemShut {NoStop}%
\bibitem [{\citenamefont {Suszalski}\ and\ \citenamefont
  {Zakrzewski}(2016)}]{SuszalskiZakrzewski2016}%
  \BibitemOpen
  \bibfield  {author} {\bibinfo {author} {\bibfnamefont {D.}~\bibnamefont
  {Suszalski}}\ and\ \bibinfo {author} {\bibfnamefont {J.}~\bibnamefont
  {Zakrzewski}},\ }\bibfield  {title} {\enquote {\bibinfo {title} {Different
  lattice geometries with synthetic dimension},}\ }\href {\doibase
  10.1103/PhysRevA.94.033602} {\bibfield  {journal} {\bibinfo  {journal} {Phys.
  Rev. A}\ }\textbf {\bibinfo {volume} {94}},\ \bibinfo {pages} {033602}
  (\bibinfo {year} {2016})}\BibitemShut {NoStop}%
\bibitem [{\citenamefont {Anisimovas}\ \emph {et~al.}(2016)\citenamefont
  {Anisimovas}, \citenamefont {Ra{\v{c}}i{\=u}nas}, \citenamefont
  {Str{\"a}ter}, \citenamefont {Eckardt}, \citenamefont {Spielman},\ and\
  \citenamefont {Juzeli{\=u}nas}}]{AnisimovasEtAl2016}%
  \BibitemOpen
  \bibfield  {author} {\bibinfo {author} {\bibfnamefont {E.}~\bibnamefont
  {Anisimovas}}, \bibinfo {author} {\bibfnamefont {M.}~\bibnamefont
  {Ra{\v{c}}i{\=u}nas}}, \bibinfo {author} {\bibfnamefont {C.}~\bibnamefont
  {Str{\"a}ter}}, \bibinfo {author} {\bibfnamefont {A.}~\bibnamefont
  {Eckardt}}, \bibinfo {author} {\bibfnamefont {I.~B.}\ \bibnamefont
  {Spielman}}, \ and\ \bibinfo {author} {\bibfnamefont {G.}~\bibnamefont
  {Juzeli{\=u}nas}},\ }\bibfield  {title} {\enquote {\bibinfo {title}
  {Semisynthetic zigzag optical lattice for ultracold bosons},}\ }\href
  {\doibase 10.1103/PhysRevA.94.063632} {\bibfield  {journal} {\bibinfo
  {journal} {Phys. Rev. A}\ }\textbf {\bibinfo {volume} {94}},\ \bibinfo
  {pages} {063632} (\bibinfo {year} {2016})}\BibitemShut {NoStop}%
\bibitem [{\citenamefont {Berry}(1984)}]{Berry1984}%
  \BibitemOpen
  \bibfield  {author} {\bibinfo {author} {\bibfnamefont {M.~V.}\ \bibnamefont
  {Berry}},\ }\bibfield  {title} {\enquote {\bibinfo {title} {Quantal phase
  factors accompanying adiabatic changes},}\ }\href {\doibase
  10.1098/rspa.1984.0023} {\bibfield  {journal} {\bibinfo  {journal} {Proc. R.
  Soc. A}\ }\textbf {\bibinfo {volume} {392}},\ \bibinfo {pages} {45--57}
  (\bibinfo {year} {1984})}\BibitemShut {NoStop}%
\bibitem [{\citenamefont {Freimuth}\ \emph {et~al.}(2013)\citenamefont
  {Freimuth}, \citenamefont {Bamler}, \citenamefont {Mokrousov},\ and\
  \citenamefont {Rosch}}]{FreimuthEtAl2013}%
  \BibitemOpen
  \bibfield  {author} {\bibinfo {author} {\bibfnamefont {F.}~\bibnamefont
  {Freimuth}}, \bibinfo {author} {\bibfnamefont {R.}~\bibnamefont {Bamler}},
  \bibinfo {author} {\bibfnamefont {Y.}~\bibnamefont {Mokrousov}}, \ and\
  \bibinfo {author} {\bibfnamefont {A.}~\bibnamefont {Rosch}},\ }\bibfield
  {title} {\enquote {\bibinfo {title} {Phase-space {B}erry phases in chiral
  magnets: {D}zyaloshinskii-{M}oriya interaction and the charge of
  skyrmions},}\ }\href {\doibase 10.1103/PhysRevB.88.214409} {\bibfield
  {journal} {\bibinfo  {journal} {Phys. Rev. B}\ }\textbf {\bibinfo {volume}
  {88}},\ \bibinfo {pages} {214409} (\bibinfo {year} {2013})}\BibitemShut
  {NoStop}%
\bibitem [{\citenamefont {Bamler}(2016)}]{Bamler2016}%
  \BibitemOpen
  \bibfield  {author} {\bibinfo {author} {\bibfnamefont {R.}~\bibnamefont
  {Bamler}},\ }\emph {\bibinfo {title} {{Phase-Space} {Berry} {Phases} in
  {Chiral} {Magnets}: {Skyrmion} {Charge}, {Hall} {Effect}, and {Dynamics} of
  {Magnetic} {Skyrmions}}},\ \href
  {http://www.thp.uni-koeln.de/~rbamler/publications/thesis-robert-bamler-2016.pdf}
  {Ph.D. thesis},\ \bibinfo  {school} {University of Cologne} (\bibinfo {year}
  {2016})\BibitemShut {NoStop}%
\bibitem [{\citenamefont {Hanneken}\ \emph {et~al.}(2015)\citenamefont
  {Hanneken}, \citenamefont {Otte}, \citenamefont {Kubetzka}, \citenamefont
  {Dup{\'e}}, \citenamefont {Romming}, \citenamefont {Von~Bergmann},
  \citenamefont {Wiesendanger},\ and\ \citenamefont
  {Heinze}}]{HannekenEtAl2015}%
  \BibitemOpen
  \bibfield  {author} {\bibinfo {author} {\bibfnamefont {C.}~\bibnamefont
  {Hanneken}}, \bibinfo {author} {\bibfnamefont {F.}~\bibnamefont {Otte}},
  \bibinfo {author} {\bibfnamefont {A.}~\bibnamefont {Kubetzka}}, \bibinfo
  {author} {\bibfnamefont {B.}~\bibnamefont {Dup{\'e}}}, \bibinfo {author}
  {\bibfnamefont {N.}~\bibnamefont {Romming}}, \bibinfo {author} {\bibfnamefont
  {K.}~\bibnamefont {Von~Bergmann}}, \bibinfo {author} {\bibfnamefont
  {R.}~\bibnamefont {Wiesendanger}}, \ and\ \bibinfo {author} {\bibfnamefont
  {S.}~\bibnamefont {Heinze}},\ }\bibfield  {title} {\enquote {\bibinfo {title}
  {{Electrical} detection of magnetic skyrmions by tunnelling non-collinear
  magnetoresistance},}\ }\href {\doibase 10.1038/nnano.2015.218} {\bibfield
  {journal} {\bibinfo  {journal} {Nat. Nanotechnol.}\ }\textbf {\bibinfo
  {volume} {10}},\ \bibinfo {pages} {1039--1042} (\bibinfo {year}
  {2015})}\BibitemShut {NoStop}%
\bibitem [{\citenamefont {Hamamoto}\ \emph {et~al.}(2016)\citenamefont
  {Hamamoto}, \citenamefont {Ezawa},\ and\ \citenamefont
  {Nagaosa}}]{HamamotoEtAl2016}%
  \BibitemOpen
  \bibfield  {author} {\bibinfo {author} {\bibfnamefont {K.}~\bibnamefont
  {Hamamoto}}, \bibinfo {author} {\bibfnamefont {M.}~\bibnamefont {Ezawa}}, \
  and\ \bibinfo {author} {\bibfnamefont {N.}~\bibnamefont {Nagaosa}},\
  }\bibfield  {title} {\enquote {\bibinfo {title} {Purely electrical detection
  of a skyrmion in constricted geometry},}\ }\href {\doibase 10.1063/1.4943949}
  {\bibfield  {journal} {\bibinfo  {journal} {Appl. Phys. Lett.}\ }\textbf
  {\bibinfo {volume} {108}},\ \bibinfo {pages} {112401} (\bibinfo {year}
  {2016})}\BibitemShut {NoStop}%
\bibitem [{\citenamefont {Sundaram}\ and\ \citenamefont
  {Niu}(1999)}]{SundaramNiu1999}%
  \BibitemOpen
  \bibfield  {author} {\bibinfo {author} {\bibfnamefont {G.}~\bibnamefont
  {Sundaram}}\ and\ \bibinfo {author} {\bibfnamefont {Q.}~\bibnamefont {Niu}},\
  }\bibfield  {title} {\enquote {\bibinfo {title} {Wave-packet dynamics in
  slowly perturbed crystals: {G}radient corrections and {B}erry-phase
  effects},}\ }\href {\doibase 10.1103/PhysRevB.59.14915} {\bibfield  {journal}
  {\bibinfo  {journal} {Phys. Rev. B}\ }\textbf {\bibinfo {volume} {59}},\
  \bibinfo {pages} {14915--14925} (\bibinfo {year} {1999})}\BibitemShut
  {NoStop}%
\bibitem [{\citenamefont {Shindou}\ and\ \citenamefont
  {Imura}(2005)}]{ShindouImura2005}%
  \BibitemOpen
  \bibfield  {author} {\bibinfo {author} {\bibfnamefont {R.}~\bibnamefont
  {Shindou}}\ and\ \bibinfo {author} {\bibfnamefont {K.-I.}\ \bibnamefont
  {Imura}},\ }\bibfield  {title} {\enquote {\bibinfo {title} {Noncommutative
  geometry and non-abelian {Berry} phase in the wave-packet dynamics of {Bloch}
  electrons},}\ }\href {\doibase 10.1016/j.nuclphysb.2005.05.019} {\bibfield
  {journal} {\bibinfo  {journal} {Nucl. Phys. B}\ }\textbf {\bibinfo {volume}
  {720}},\ \bibinfo {pages} {399--435} (\bibinfo {year} {2005})}\BibitemShut
  {NoStop}%
\bibitem [{\citenamefont {Gosselin}\ \emph {et~al.}(2006)\citenamefont
  {Gosselin}, \citenamefont {M{\'e}nas}, \citenamefont {B{\'e}rard},\ and\
  \citenamefont {Mohrbach}}]{Gosselin2006}%
  \BibitemOpen
  \bibfield  {author} {\bibinfo {author} {\bibfnamefont {P.}~\bibnamefont
  {Gosselin}}, \bibinfo {author} {\bibfnamefont {F.}~\bibnamefont {M{\'e}nas}},
  \bibinfo {author} {\bibfnamefont {A.}~\bibnamefont {B{\'e}rard}}, \ and\
  \bibinfo {author} {\bibfnamefont {H.}~\bibnamefont {Mohrbach}},\ }\bibfield
  {title} {\enquote {\bibinfo {title} {Semiclassical dynamics of electrons in
  magnetic {Bloch} bands: {A} {Hamiltonian} approach},}\ }\href
  {http://stacks.iop.org/0295-5075/76/i=4/a=651} {\bibfield  {journal}
  {\bibinfo  {journal} {Europhys. Lett.}\ }\textbf {\bibinfo {volume} {76}},\
  \bibinfo {pages} {651} (\bibinfo {year} {2006})}\BibitemShut {NoStop}%
\bibitem [{\citenamefont {Gosselin}\ \emph {et~al.}(2007)\citenamefont
  {Gosselin}, \citenamefont {B{\'e}rard},\ and\ \citenamefont
  {Mohrbach}}]{Gosselin2007}%
  \BibitemOpen
  \bibfield  {author} {\bibinfo {author} {\bibfnamefont {P.}~\bibnamefont
  {Gosselin}}, \bibinfo {author} {\bibfnamefont {A.}~\bibnamefont
  {B{\'e}rard}}, \ and\ \bibinfo {author} {\bibfnamefont {H.}~\bibnamefont
  {Mohrbach}},\ }\bibfield  {title} {\enquote {\bibinfo {title} {Semiclassical
  diagonalization of quantum {Hamiltonian} and equations of motion with {Berry}
  phase corrections},}\ }\href {\doibase 10.1140/epjb/e2007-00212-6} {\bibfield
   {journal} {\bibinfo  {journal} {EPJ B}\ }\textbf {\bibinfo {volume} {58}},\
  \bibinfo {pages} {137--148} (\bibinfo {year} {2007})}\BibitemShut {NoStop}%
\bibitem [{\citenamefont {Gosselin}\ \emph
  {et~al.}(2008{\natexlab{a}})\citenamefont {Gosselin}, \citenamefont
  {Boumrar},\ and\ \citenamefont {Mohrbach}}]{Gosselin2008a}%
  \BibitemOpen
  \bibfield  {author} {\bibinfo {author} {\bibfnamefont {P.}~\bibnamefont
  {Gosselin}}, \bibinfo {author} {\bibfnamefont {H.}~\bibnamefont {Boumrar}}, \
  and\ \bibinfo {author} {\bibfnamefont {H.}~\bibnamefont {Mohrbach}},\
  }\bibfield  {title} {\enquote {\bibinfo {title} {Semiclassical quantization
  of electrons in magnetic fields: the generalized {Peierls} substitution},}\
  }\href {\doibase 10.1209/0295-5075/84/50002} {\bibfield  {journal} {\bibinfo
  {journal} {EPL}\ }\textbf {\bibinfo {volume} {84}},\ \bibinfo {pages} {50002}
  (\bibinfo {year} {2008}{\natexlab{a}})}\BibitemShut {NoStop}%
\bibitem [{\citenamefont {Gosselin}\ \emph
  {et~al.}(2008{\natexlab{b}})\citenamefont {Gosselin}, \citenamefont
  {Hanssen},\ and\ \citenamefont {Mohrbach}}]{Gosselin2008b}%
  \BibitemOpen
  \bibfield  {author} {\bibinfo {author} {\bibfnamefont {P.}~\bibnamefont
  {Gosselin}}, \bibinfo {author} {\bibfnamefont {J.}~\bibnamefont {Hanssen}}, \
  and\ \bibinfo {author} {\bibfnamefont {H.}~\bibnamefont {Mohrbach}},\
  }\bibfield  {title} {\enquote {\bibinfo {title} {Recursive diagonalization of
  quantum hamiltonians to all orders in $\ensuremath{\hbar}$},}\ }\href
  {\doibase 10.1103/PhysRevD.77.085008} {\bibfield  {journal} {\bibinfo
  {journal} {Phys. Rev. D}\ }\textbf {\bibinfo {volume} {77}},\ \bibinfo
  {pages} {085008} (\bibinfo {year} {2008}{\natexlab{b}})}\BibitemShut
  {NoStop}%
\bibitem [{\citenamefont {Gosselin}\ and\ \citenamefont
  {Mohrbach}(2009)}]{Gosselin2009}%
  \BibitemOpen
  \bibfield  {author} {\bibinfo {author} {\bibfnamefont {P.}~\bibnamefont
  {Gosselin}}\ and\ \bibinfo {author} {\bibfnamefont {H.}~\bibnamefont
  {Mohrbach}},\ }\bibfield  {title} {\enquote {\bibinfo {title} {Diagonal
  representation for a generic matrix valued quantum {Hamiltonian}},}\ }\href
  {\doibase 10.1140/epjc/s10052-009-1155-3} {\bibfield  {journal} {\bibinfo
  {journal} {Eur. Phys. J. C}\ }\textbf {\bibinfo {volume} {64}},\ \bibinfo
  {pages} {495--527} (\bibinfo {year} {2009})}\BibitemShut {NoStop}%
\bibitem [{\citenamefont {Wong}\ and\ \citenamefont
  {Tserkovnyak}(2011)}]{WongTserkovnyak2011}%
  \BibitemOpen
  \bibfield  {author} {\bibinfo {author} {\bibfnamefont {C.~H.}\ \bibnamefont
  {Wong}}\ and\ \bibinfo {author} {\bibfnamefont {Y.}~\bibnamefont
  {Tserkovnyak}},\ }\bibfield  {title} {\enquote {\bibinfo {title} {Quantum
  kinetic equation in phase-space textured multiband systems},}\ }\href
  {\doibase 10.1103/PhysRevB.84.115209} {\bibfield  {journal} {\bibinfo
  {journal} {Phys. Rev. B}\ }\textbf {\bibinfo {volume} {84}},\ \bibinfo
  {pages} {115209} (\bibinfo {year} {2011})}\BibitemShut {NoStop}%
\bibitem [{\citenamefont {Su}\ \emph {et~al.}(2015)\citenamefont {Su},
  \citenamefont {Gou}, \citenamefont {Liu}, \citenamefont {Spielman},
  \citenamefont {Santos}, \citenamefont {Acus}, \citenamefont {Mekys},
  \citenamefont {Ruseckas},\ and\ \citenamefont {Juzeli{\=u}nas}}]{Su2015}%
  \BibitemOpen
  \bibfield  {author} {\bibinfo {author} {\bibfnamefont {S.-W.}\ \bibnamefont
  {Su}}, \bibinfo {author} {\bibfnamefont {S.-C.}\ \bibnamefont {Gou}},
  \bibinfo {author} {\bibfnamefont {I.-K.}\ \bibnamefont {Liu}}, \bibinfo
  {author} {\bibfnamefont {I.~B.}\ \bibnamefont {Spielman}}, \bibinfo {author}
  {\bibfnamefont {L.}~\bibnamefont {Santos}}, \bibinfo {author} {\bibfnamefont
  {A.}~\bibnamefont {Acus}}, \bibinfo {author} {\bibfnamefont {A.}~\bibnamefont
  {Mekys}}, \bibinfo {author} {\bibfnamefont {J.}~\bibnamefont {Ruseckas}}, \
  and\ \bibinfo {author} {\bibfnamefont {G.}~\bibnamefont {Juzeli{\=u}nas}},\
  }\bibfield  {title} {\enquote {\bibinfo {title} {Position-dependent
  spin–orbit coupling for ultracold atoms},}\ }\href {\doibase
  10.1088/1367-2630/17/3/033045} {\bibfield  {journal} {\bibinfo  {journal}
  {New J. Phys.}\ }\textbf {\bibinfo {volume} {17}},\ \bibinfo {pages} {033045}
  (\bibinfo {year} {2015})}\BibitemShut {NoStop}%
\bibitem [{\citenamefont {Zhang}\ \emph {et~al.}(2016)\citenamefont {Zhang},
  \citenamefont {Yu}, \citenamefont {Ng}, \citenamefont {Zhang}, \citenamefont
  {Pitaevskii},\ and\ \citenamefont {Stringari}}]{ZhangEtAl2016}%
  \BibitemOpen
  \bibfield  {author} {\bibinfo {author} {\bibfnamefont {Y.-C.}\ \bibnamefont
  {Zhang}}, \bibinfo {author} {\bibfnamefont {Z.-Q.}\ \bibnamefont {Yu}},
  \bibinfo {author} {\bibfnamefont {T.~K.}\ \bibnamefont {Ng}}, \bibinfo
  {author} {\bibfnamefont {S.}~\bibnamefont {Zhang}}, \bibinfo {author}
  {\bibfnamefont {L.}~\bibnamefont {Pitaevskii}}, \ and\ \bibinfo {author}
  {\bibfnamefont {S.}~\bibnamefont {Stringari}},\ }\bibfield  {title} {\enquote
  {\bibinfo {title} {Superfluid density of a spin-orbit-coupled {Bose} gas},}\
  }\href {\doibase 10.1103/PhysRevA.94.033635} {\bibfield  {journal} {\bibinfo
  {journal} {Phys. Rev. A}\ }\textbf {\bibinfo {volume} {94}},\ \bibinfo
  {pages} {033635} (\bibinfo {year} {2016})}\BibitemShut {NoStop}%
\bibitem [{\citenamefont {Seiberg}\ and\ \citenamefont
  {Witten}(1999)}]{SeibergWitten1999}%
  \BibitemOpen
  \bibfield  {author} {\bibinfo {author} {\bibfnamefont {N.}~\bibnamefont
  {Seiberg}}\ and\ \bibinfo {author} {\bibfnamefont {E.}~\bibnamefont
  {Witten}},\ }\bibfield  {title} {\enquote {\bibinfo {title} {String theory
  and noncommutative geometry},}\ }\href {\doibase
  10.1088/1126-6708/1999/09/032} {\bibfield  {journal} {\bibinfo  {journal} {J.
  High Energy Phys.}\ }\textbf {\bibinfo {volume} {1999}},\ \bibinfo {pages}
  {032} (\bibinfo {year} {1999})}\BibitemShut {NoStop}%
\bibitem [{\citenamefont {Douglas}\ and\ \citenamefont
  {Nekrasov}(2001)}]{DouglasNekrasov2001}%
  \BibitemOpen
  \bibfield  {author} {\bibinfo {author} {\bibfnamefont {M.~R.}\ \bibnamefont
  {Douglas}}\ and\ \bibinfo {author} {\bibfnamefont {N.~A.}\ \bibnamefont
  {Nekrasov}},\ }\bibfield  {title} {\enquote {\bibinfo {title}
  {{Noncommutative} field theory},}\ }\href {\doibase
  10.1103/RevModPhys.73.977} {\bibfield  {journal} {\bibinfo  {journal} {Rev.
  Mod. Phys.}\ }\textbf {\bibinfo {volume} {73}},\ \bibinfo {pages} {977--1029}
  (\bibinfo {year} {2001})}\BibitemShut {NoStop}%
\bibitem [{\citenamefont {Merkl}\ \emph {et~al.}(2008)\citenamefont {Merkl},
  \citenamefont {Zimmer}, \citenamefont {Juzeli\={u}nas},\ and\ \citenamefont
  {\"Ohberg}}]{Merkl08}%
  \BibitemOpen
  \bibfield  {author} {\bibinfo {author} {\bibfnamefont {M.}~\bibnamefont
  {Merkl}}, \bibinfo {author} {\bibfnamefont {F.~E.}\ \bibnamefont {Zimmer}},
  \bibinfo {author} {\bibfnamefont {G.}~\bibnamefont {Juzeli\={u}nas}}, \ and\
  \bibinfo {author} {\bibfnamefont {P.}~\bibnamefont {\"Ohberg}},\ }\bibfield
  {title} {\enquote {\bibinfo {title} {{Atomic} {Zitterbewegung}},}\ }\href
  {\doibase 10.1209/0295-5075/83/54002} {\bibfield  {journal} {\bibinfo
  {journal} {EPL}\ }\textbf {\bibinfo {volume} {83}},\ \bibinfo {pages} {54002}
  (\bibinfo {year} {2008})}\BibitemShut {NoStop}%
\bibitem [{\citenamefont {LeBlanc}\ \emph {et~al.}(2013)\citenamefont
  {LeBlanc}, \citenamefont {Beeler}, \citenamefont {Jimenez-Garcia},
  \citenamefont {Perry}, \citenamefont {Sugawa}, \citenamefont {Williams},\
  and\ \citenamefont {Spielman}}]{LeBlanc2013}%
  \BibitemOpen
  \bibfield  {author} {\bibinfo {author} {\bibfnamefont {L.~J.}\ \bibnamefont
  {LeBlanc}}, \bibinfo {author} {\bibfnamefont {M.~C.}\ \bibnamefont {Beeler}},
  \bibinfo {author} {\bibfnamefont {K.}~\bibnamefont {Jimenez-Garcia}},
  \bibinfo {author} {\bibfnamefont {A.~R.}\ \bibnamefont {Perry}}, \bibinfo
  {author} {\bibfnamefont {S.}~\bibnamefont {Sugawa}}, \bibinfo {author}
  {\bibfnamefont {R.~A.}\ \bibnamefont {Williams}}, \ and\ \bibinfo {author}
  {\bibfnamefont {I.~B.}\ \bibnamefont {Spielman}},\ }\bibfield  {title}
  {\enquote {\bibinfo {title} {Direct observation of {Zitterbewegung} in a
  {Bose}-{Einstein} condensate},}\ }\href {\doibase
  10.1088/1367-2630/15/7/073011} {\bibfield  {journal} {\bibinfo  {journal}
  {New. J. Phys.}\ }\textbf {\bibinfo {volume} {15}},\ \bibinfo {pages}
  {073011} (\bibinfo {year} {2013})}\BibitemShut {NoStop}%
\bibitem [{\citenamefont {Winkler}\ \emph {et~al.}(2007)\citenamefont
  {Winkler}, \citenamefont {Z\"ulicke},\ and\ \citenamefont
  {Bolte}}]{WinklerEtAl2007}%
  \BibitemOpen
  \bibfield  {author} {\bibinfo {author} {\bibfnamefont {R.}~\bibnamefont
  {Winkler}}, \bibinfo {author} {\bibfnamefont {U.}~\bibnamefont {Z\"ulicke}},
  \ and\ \bibinfo {author} {\bibfnamefont {J.}~\bibnamefont {Bolte}},\
  }\bibfield  {title} {\enquote {\bibinfo {title} {Oscillatory multiband
  dynamics of free particles: The ubiquity of zitterbewegung effects},}\ }\href
  {\doibase 10.1103/PhysRevB.75.205314} {\bibfield  {journal} {\bibinfo
  {journal} {Phys. Rev. B}\ }\textbf {\bibinfo {volume} {75}},\ \bibinfo
  {pages} {205314} (\bibinfo {year} {2007})}\BibitemShut {NoStop}%
\bibitem [{\citenamefont {Volovik}(1987)}]{Volovik1987}%
  \BibitemOpen
  \bibfield  {author} {\bibinfo {author} {\bibfnamefont {G.~E.}\ \bibnamefont
  {Volovik}},\ }\bibfield  {title} {\enquote {\bibinfo {title} {Linear momentum
  in ferromagnets},}\ }\href {http://stacks.iop.org/0022-3719/20/i=7/a=003}
  {\bibfield  {journal} {\bibinfo  {journal} {J. Phys. C: Solid State Phys.}\
  }\textbf {\bibinfo {volume} {20}},\ \bibinfo {pages} {L83} (\bibinfo {year}
  {1987})}\BibitemShut {NoStop}%
\bibitem [{\citenamefont {Lin}\ \emph {et~al.}(2009)\citenamefont {Lin},
  \citenamefont {Compton}, \citenamefont {{Jim{\'e}nez-Garc{\'{\i}}a}},
  \citenamefont {Porto},\ and\ \citenamefont {Spielman}}]{LinEtAl2009a}%
  \BibitemOpen
  \bibfield  {author} {\bibinfo {author} {\bibfnamefont {Y.-J.}\ \bibnamefont
  {Lin}}, \bibinfo {author} {\bibfnamefont {R.~L.}\ \bibnamefont {Compton}},
  \bibinfo {author} {\bibfnamefont {K.}~\bibnamefont
  {{Jim{\'e}nez-Garc{\'{\i}}a}}}, \bibinfo {author} {\bibfnamefont {J.~V.}\
  \bibnamefont {Porto}}, \ and\ \bibinfo {author} {\bibfnamefont {I.~B.}\
  \bibnamefont {Spielman}},\ }\bibfield  {title} {\enquote {\bibinfo {title}
  {Synthetic magnetic fields for ultracold neutral atoms},}\ }\href {\doibase
  10.1038/nature08609} {\bibfield  {journal} {\bibinfo  {journal} {Nature}\
  }\textbf {\bibinfo {volume} {462}},\ \bibinfo {pages} {628--632} (\bibinfo
  {year} {2009})}\BibitemShut {NoStop}%
\bibitem [{\citenamefont {Beeler}\ \emph {et~al.}(2013)\citenamefont {Beeler},
  \citenamefont {Williams}, \citenamefont {Jimenez-Garcia}, \citenamefont
  {LeBlanc}, \citenamefont {Perry},\ and\ \citenamefont
  {Spielman}}]{BeelerEtAl2013}%
  \BibitemOpen
  \bibfield  {author} {\bibinfo {author} {\bibfnamefont {M.~C.}\ \bibnamefont
  {Beeler}}, \bibinfo {author} {\bibfnamefont {R.~A.}\ \bibnamefont
  {Williams}}, \bibinfo {author} {\bibfnamefont {K.}~\bibnamefont
  {Jimenez-Garcia}}, \bibinfo {author} {\bibfnamefont {L.~J.}\ \bibnamefont
  {LeBlanc}}, \bibinfo {author} {\bibfnamefont {A.~R.}\ \bibnamefont {Perry}},
  \ and\ \bibinfo {author} {\bibfnamefont {I.~B.}\ \bibnamefont {Spielman}},\
  }\bibfield  {title} {\enquote {\bibinfo {title} {{The} spin {Hall} effect in
  a quantum gas},}\ }\href {\doibase 10.1038/nature12185} {\bibfield  {journal}
  {\bibinfo  {journal} {Nature}\ }\textbf {\bibinfo {volume} {498}},\ \bibinfo
  {pages} {201--204} (\bibinfo {year} {2013})}\BibitemShut {NoStop}%
\bibitem [{\citenamefont {Gerlach}\ and\ \citenamefont
  {Stern}(1922)}]{GerlachStern1922}%
  \BibitemOpen
  \bibfield  {author} {\bibinfo {author} {\bibfnamefont {W.}~\bibnamefont
  {Gerlach}}\ and\ \bibinfo {author} {\bibfnamefont {O.}~\bibnamefont
  {Stern}},\ }\bibfield  {title} {\enquote {\bibinfo {title} {{Der}
  experimentelle {Nachweis} der {Richtungsquantelung} im {Magnetfeld}},}\
  }\href {\doibase 10.1007/BF01326983} {\bibfield  {journal} {\bibinfo
  {journal} {Zeitschrift f{\"u}r Physik}\ }\textbf {\bibinfo {volume} {9}},\
  \bibinfo {pages} {349--352} (\bibinfo {year} {1922})}\BibitemShut {NoStop}%
\bibitem [{\citenamefont {Sakurai}(1994)}]{Sakurai1994}%
  \BibitemOpen
  \bibfield  {author} {\bibinfo {author} {\bibfnamefont {J.~J.}\ \bibnamefont
  {Sakurai}},\ }\href@noop {} {\emph {\bibinfo {title} {{Modern} {Quantum}
  {Mechanics}}}},\ \bibinfo {edition} {{R}evised}\ ed.\ (\bibinfo  {publisher}
  {{Addison}-{Wesley} {Publishing} {Company}},\ \bibinfo {year}
  {1994})\BibitemShut {NoStop}%
\bibitem [{\citenamefont {Price}\ \emph {et~al.}(2014)\citenamefont {Price},
  \citenamefont {Ozawa},\ and\ \citenamefont {Carusotto}}]{PriceEtAl2014}%
  \BibitemOpen
  \bibfield  {author} {\bibinfo {author} {\bibfnamefont {H.~M.}\ \bibnamefont
  {Price}}, \bibinfo {author} {\bibfnamefont {T.}~\bibnamefont {Ozawa}}, \ and\
  \bibinfo {author} {\bibfnamefont {I.}~\bibnamefont {Carusotto}},\ }\bibfield
  {title} {\enquote {\bibinfo {title} {{Quantum} {Mechanics} with a
  {Momentum}-{Space} {Artificial} {Magnetic} {Field}},}\ }\href {\doibase
  10.1103/PhysRevLett.113.190403} {\bibfield  {journal} {\bibinfo  {journal}
  {Phys. Rev. Lett.}\ }\textbf {\bibinfo {volume} {113}},\ \bibinfo {pages}
  {190403} (\bibinfo {year} {2014})}\BibitemShut {NoStop}%
\bibitem [{\citenamefont {Price}\ \emph
  {et~al.}(2015{\natexlab{b}})\citenamefont {Price}, \citenamefont {Ozawa},
  \citenamefont {Cooper},\ and\ \citenamefont {Carusotto}}]{PriceEtAl2015b}%
  \BibitemOpen
  \bibfield  {author} {\bibinfo {author} {\bibfnamefont {Hannah~M.}\
  \bibnamefont {Price}}, \bibinfo {author} {\bibfnamefont {Tomoki}\
  \bibnamefont {Ozawa}}, \bibinfo {author} {\bibfnamefont {Nigel~R.}\
  \bibnamefont {Cooper}}, \ and\ \bibinfo {author} {\bibfnamefont {Iacopo}\
  \bibnamefont {Carusotto}},\ }\bibfield  {title} {\enquote {\bibinfo {title}
  {{Artificial} magnetic fields in momentum space in spin-orbit-coupled
  systems},}\ }\href {\doibase 10.1103/PhysRevA.91.033606} {\bibfield
  {journal} {\bibinfo  {journal} {Phys. Rev. A}\ }\textbf {\bibinfo {volume}
  {91}},\ \bibinfo {pages} {033606} (\bibinfo {year}
  {2015}{\natexlab{b}})}\BibitemShut {NoStop}%
\bibitem [{\citenamefont {Lin}\ \emph {et~al.}(2011{\natexlab{b}})\citenamefont
  {Lin}, \citenamefont {Compton}, \citenamefont {Jimenez-Garcia}, \citenamefont
  {Phillips}, \citenamefont {Porto},\ and\ \citenamefont
  {Spielman}}]{LinEtAl2011b}%
  \BibitemOpen
  \bibfield  {author} {\bibinfo {author} {\bibfnamefont {Y.-J.}\ \bibnamefont
  {Lin}}, \bibinfo {author} {\bibfnamefont {R.~L.}\ \bibnamefont {Compton}},
  \bibinfo {author} {\bibfnamefont {K.}~\bibnamefont {Jimenez-Garcia}},
  \bibinfo {author} {\bibfnamefont {W.~D.}\ \bibnamefont {Phillips}}, \bibinfo
  {author} {\bibfnamefont {J.~V.}\ \bibnamefont {Porto}}, \ and\ \bibinfo
  {author} {\bibfnamefont {I.~B.}\ \bibnamefont {Spielman}},\ }\bibfield
  {title} {\enquote {\bibinfo {title} {A synthetic electric force acting on
  neutral atoms},}\ }\href {\doibase 10.1038/nphys1954} {\bibfield  {journal}
  {\bibinfo  {journal} {Nature Phys.}\ }\textbf {\bibinfo {volume} {7}},\
  \bibinfo {pages} {531--534} (\bibinfo {year}
  {2011}{\natexlab{b}})}\BibitemShut {NoStop}%
\bibitem [{\citenamefont {Zhang}\ \emph {et~al.}(2012)\citenamefont {Zhang},
  \citenamefont {Ji}, \citenamefont {Chen}, \citenamefont {Zhang},
  \citenamefont {Du}, \citenamefont {Yan}, \citenamefont {Pan}, \citenamefont
  {Zhao}, \citenamefont {Deng}, \citenamefont {Zhai}, \citenamefont {Chen},\
  and\ \citenamefont {Pan}}]{ZhangEtAl2012}%
  \BibitemOpen
  \bibfield  {author} {\bibinfo {author} {\bibfnamefont {J.-Y.}\ \bibnamefont
  {Zhang}}, \bibinfo {author} {\bibfnamefont {S.-C.}\ \bibnamefont {Ji}},
  \bibinfo {author} {\bibfnamefont {Z.}~\bibnamefont {Chen}}, \bibinfo {author}
  {\bibfnamefont {L.}~\bibnamefont {Zhang}}, \bibinfo {author} {\bibfnamefont
  {Z.-D.}\ \bibnamefont {Du}}, \bibinfo {author} {\bibfnamefont
  {B.}~\bibnamefont {Yan}}, \bibinfo {author} {\bibfnamefont {G.-S.}\
  \bibnamefont {Pan}}, \bibinfo {author} {\bibfnamefont {B.}~\bibnamefont
  {Zhao}}, \bibinfo {author} {\bibfnamefont {Y.-J.}\ \bibnamefont {Deng}},
  \bibinfo {author} {\bibfnamefont {H.}~\bibnamefont {Zhai}}, \bibinfo {author}
  {\bibfnamefont {S.}~\bibnamefont {Chen}}, \ and\ \bibinfo {author}
  {\bibfnamefont {J.-W.}\ \bibnamefont {Pan}},\ }\bibfield  {title} {\enquote
  {\bibinfo {title} {{Collective} {Dipole} {Oscillations} of a {Spin-Orbit}
  {Coupled} {Bose-Einstein} {Condensate}},}\ }\href {\doibase
  10.1103/PhysRevLett.109.115301} {\bibfield  {journal} {\bibinfo  {journal}
  {Phys. Rev. Lett.}\ }\textbf {\bibinfo {volume} {109}},\ \bibinfo {pages}
  {115301} (\bibinfo {year} {2012})}\BibitemShut {NoStop}%
\bibitem [{\citenamefont {Aidelsburger}\ \emph {et~al.}(2015)\citenamefont
  {Aidelsburger}, \citenamefont {Lohse}, \citenamefont {Schweizer},
  \citenamefont {Atala}, \citenamefont {Barreiro}, \citenamefont
  {Nascimb\`{e}ne}, \citenamefont {Cooper}, \citenamefont {Bloch},\ and\
  \citenamefont {Goldman}}]{AidelsburgerEtAl2015}%
  \BibitemOpen
  \bibfield  {author} {\bibinfo {author} {\bibfnamefont {M.}~\bibnamefont
  {Aidelsburger}}, \bibinfo {author} {\bibfnamefont {M.}~\bibnamefont {Lohse}},
  \bibinfo {author} {\bibfnamefont {C.}~\bibnamefont {Schweizer}}, \bibinfo
  {author} {\bibfnamefont {M.}~\bibnamefont {Atala}}, \bibinfo {author}
  {\bibfnamefont {J.~T.}\ \bibnamefont {Barreiro}}, \bibinfo {author}
  {\bibfnamefont {S.}~\bibnamefont {Nascimb\`{e}ne}}, \bibinfo {author}
  {\bibfnamefont {N.~R.}\ \bibnamefont {Cooper}}, \bibinfo {author}
  {\bibfnamefont {I.}~\bibnamefont {Bloch}}, \ and\ \bibinfo {author}
  {\bibfnamefont {N.}~\bibnamefont {Goldman}},\ }\bibfield  {title} {\enquote
  {\bibinfo {title} {Measuring the {C}hern number of {H}ofstadter bands with
  ultracold bosonic atoms},}\ }\href {\doibase 10.1038/nphys3171} {\bibfield
  {journal} {\bibinfo  {journal} {Nat. Phys.}\ }\textbf {\bibinfo {volume}
  {11}},\ \bibinfo {pages} {162--166} (\bibinfo {year} {2015})}\BibitemShut
  {NoStop}%
\bibitem [{\citenamefont {Kolkowitz}\ \emph {et~al.}(2017)\citenamefont
  {Kolkowitz}, \citenamefont {Bromley}, \citenamefont {Bothwell}, \citenamefont
  {Wall}, \citenamefont {Marti}, \citenamefont {Koller}, \citenamefont {Zhang},
  \citenamefont {Rey},\ and\ \citenamefont {Ye}}]{KolkowitzEtAl2017}%
  \BibitemOpen
  \bibfield  {author} {\bibinfo {author} {\bibfnamefont {S.}~\bibnamefont
  {Kolkowitz}}, \bibinfo {author} {\bibfnamefont {S.~L.}\ \bibnamefont
  {Bromley}}, \bibinfo {author} {\bibfnamefont {T.}~\bibnamefont {Bothwell}},
  \bibinfo {author} {\bibfnamefont {M.~L.}\ \bibnamefont {Wall}}, \bibinfo
  {author} {\bibfnamefont {G.~E.}\ \bibnamefont {Marti}}, \bibinfo {author}
  {\bibfnamefont {A.~P.}\ \bibnamefont {Koller}}, \bibinfo {author}
  {\bibfnamefont {X.}~\bibnamefont {Zhang}}, \bibinfo {author} {\bibfnamefont
  {A.~M.}\ \bibnamefont {Rey}}, \ and\ \bibinfo {author} {\bibfnamefont
  {J.}~\bibnamefont {Ye}},\ }\bibfield  {title} {\enquote {\bibinfo {title}
  {Spin--orbit-coupled fermions in an optical lattice clock},}\ }\href
  {\doibase 10.1038/nature20811} {\bibfield  {journal} {\bibinfo  {journal}
  {Nature}\ }\textbf {\bibinfo {volume} {542}},\ \bibinfo {pages} {66--70}
  (\bibinfo {year} {2017})}\BibitemShut {NoStop}%
\bibitem [{\citenamefont {Alberti}\ \emph {et~al.}(2009)\citenamefont
  {Alberti}, \citenamefont {Ivanov}, \citenamefont {Tino},\ and\ \citenamefont
  {Ferrari}}]{AlbertiEtAl2009}%
  \BibitemOpen
  \bibfield  {author} {\bibinfo {author} {\bibfnamefont {A.}~\bibnamefont
  {Alberti}}, \bibinfo {author} {\bibfnamefont {V.~V.}\ \bibnamefont {Ivanov}},
  \bibinfo {author} {\bibfnamefont {G.~M.}\ \bibnamefont {Tino}}, \ and\
  \bibinfo {author} {\bibfnamefont {G.}~\bibnamefont {Ferrari}},\ }\bibfield
  {title} {\enquote {\bibinfo {title} {Engineering the quantum transport of
  atomic wavefunctions over macroscopic distances},}\ }\href {\doibase
  10.1038/nphys1310} {\bibfield  {journal} {\bibinfo  {journal} {Nat. Phys.}\
  }\textbf {\bibinfo {volume} {5}},\ \bibinfo {pages} {547--550} (\bibinfo
  {year} {2009})}\BibitemShut {NoStop}%
\bibitem [{\citenamefont {Alberti}\ \emph {et~al.}(2010)\citenamefont
  {Alberti}, \citenamefont {Ferrari}, \citenamefont {Ivanov}, \citenamefont
  {Chiofalo},\ and\ \citenamefont {Tino}}]{AlbertiEtAl2010}%
  \BibitemOpen
  \bibfield  {author} {\bibinfo {author} {\bibfnamefont {A.}~\bibnamefont
  {Alberti}}, \bibinfo {author} {\bibfnamefont {G.}~\bibnamefont {Ferrari}},
  \bibinfo {author} {\bibfnamefont {V.~V.}\ \bibnamefont {Ivanov}}, \bibinfo
  {author} {\bibfnamefont {M.~L.}\ \bibnamefont {Chiofalo}}, \ and\ \bibinfo
  {author} {\bibfnamefont {G.~M.}\ \bibnamefont {Tino}},\ }\bibfield  {title}
  {\enquote {\bibinfo {title} {Atomic wave packets in amplitude-modulated
  vertical optical lattices},}\ }\href {\doibase 10.1088/1367-2630/12/6/065037}
  {\bibfield  {journal} {\bibinfo  {journal} {New J. Phys.}\ }\textbf {\bibinfo
  {volume} {12}},\ \bibinfo {pages} {065037} (\bibinfo {year}
  {2010})}\BibitemShut {NoStop}%
\bibitem [{\citenamefont {Haller}\ \emph {et~al.}(2010)\citenamefont {Haller},
  \citenamefont {Hart}, \citenamefont {Mark}, \citenamefont {Danzl},
  \citenamefont {Reichs{\"o}llner},\ and\ \citenamefont
  {N{\"a}gerl}}]{HallerEtAl2010}%
  \BibitemOpen
  \bibfield  {author} {\bibinfo {author} {\bibfnamefont {E.}~\bibnamefont
  {Haller}}, \bibinfo {author} {\bibfnamefont {R.}~\bibnamefont {Hart}},
  \bibinfo {author} {\bibfnamefont {M.~J.}\ \bibnamefont {Mark}}, \bibinfo
  {author} {\bibfnamefont {J.~G.}\ \bibnamefont {Danzl}}, \bibinfo {author}
  {\bibfnamefont {L.}~\bibnamefont {Reichs{\"o}llner}}, \ and\ \bibinfo
  {author} {\bibfnamefont {H.-C.}\ \bibnamefont {N{\"a}gerl}},\ }\bibfield
  {title} {\enquote {\bibinfo {title} {Inducing transport in a dissipation-free
  lattice with super {B}loch oscillations},}\ }\href {\doibase
  10.1103/PhysRevLett.104.200403} {\bibfield  {journal} {\bibinfo  {journal}
  {Phys. Rev. Lett.}\ }\textbf {\bibinfo {volume} {104}},\ \bibinfo {pages}
  {200403} (\bibinfo {year} {2010})}\BibitemShut {NoStop}%
\end{thebibliography}%

\end{document}